\documentclass[reprint,aps,pre,onecolumn,12pt,amsmath,floatfix,superscriptaddress]{revtex4-1}
\usepackage[utf8]{inputenc}
\usepackage{graphicx}
\usepackage{amsmath,amsfonts,amssymb}
\usepackage[usenames,dvipsnames]{color}
\usepackage[normalem]{ulem}

\begin{document}
\title{Morphological transitions of sliding drops -- Dynamics and bifurcations}
\author{Sebastian Engelnkemper}
\email[]{engelnkemper@uni-muenster.de}
\affiliation{Institute for Theoretical Physics, University of M\"unster, Wilhelm-Klemm-Strasse\ 9, D-48149 M\"unster, Germany}
\author{Markus Wilczek}
\affiliation{Institute for Theoretical Physics, University of M\"unster, Wilhelm-Klemm-Strasse\ 9, D-48149 M\"unster, Germany}
\affiliation{Center for Nonlinear Science (CeNoS), University of M\"unster, Corrensstrasse\ 2, D-48149 M\"unster, Germany}
\author{Svetlana V. Gurevich}
\author{Uwe Thiele}
\email[]{u.thiele@uni-muenster.de}
\affiliation{Institute for Theoretical Physics, University of M\"unster, Wilhelm-Klemm-Strasse\ 9, D-48149 M\"unster, Germany}
\affiliation{Center for Nonlinear Science (CeNoS), University of M\"unster, Corrensstrasse\ 2, D-48149 M\"unster, Germany}
\affiliation{Center for Multiscale Theory and Computation (CMTC), University of M\"unster, Corrensstrasse\ 40, D-48149 M\"unster, Germany}
\begin{abstract}
We study fully three-dimensional droplets that slide down an incline by employing a thin-film equation that accounts for capillarity, wettability, and a lateral driving force in small-gradient (or long-wave) approximation. In particular, we focus on qualitative changes in the morphology and behavior of stationary sliding drops. We employ the inclination angle of the substrate as control parameter and use continuation techniques to analyze for several fixed droplet sizes the bifurcation diagram of stationary droplets, their linear stability, and relevant eigenmodes. The obtained predictions on existence ranges and instabilities are tested via direct numerical simulations that are also used to investigate a branch of time-periodic behavior (corresponding to repeated breakup-coalescence cycles, where the breakup is also denoted as \textsl{pearling}) which emerges at a global instability, the related hysteresis in behavior, and a period-doubling cascade. The non trivial oscillatory behavior close to a Hopf bifurcation of drops with a finite-length tail is also studied. Finally, it is shown that the main features of the bifurcation diagram follow scaling laws over several decades of the droplet size.
\end{abstract}

\maketitle

\section{Introduction}
It is known from many experiments and everyday experience that drops on a smooth homogeneous solid substrate slide along the substrate when lateral driving forces are applied. For heterogeneous substrates the driving force first needs to overcome pinning influences of the heterogeneities. Physical systems where individual or many sliding drops can be observed are, e.g., rain drops on a train window driven by the shear flow of the outside airstream, drops sliding or rolling down an incline due to gravitational force \cite{HuMa1991l,RiQu1999el,Podgorski2001Corners} or drops moving outwards on a spinning disk due to the centrifugal force. The motion of droplets may also be induced by chemical or thermal gradients along the substrate \cite{Gree1978jfm,Broc1989l,ChWh1992s}. In all these settings, control parameters, such as, droplet volume, strength of the driving force, material properties of the liquid (viscosity, surface tension) and interaction properties of the combination liquid-substrate (equilibrium contact angle) determine the velocity and shape of the moving droplets.

Here, we mainly focus on sliding droplets of partially wetting liquids on an incline, i.e., droplets that form at equilibrium finite contact angles with the substrate and slide down under the influence of gravity. Both, experimental \cite{Podgorski2001Corners,Kim2002Sliding,legrand2005shape,snoeijer2007cornered} and theoretical \cite{BeCP2001crassib,TVNB2001pre,Thiele2002Sliding,SLWF2002crp,amar2003transition,limat2004three,Schwartz2005Shapes,benilov2015thin,XDD2016PF} studies, analyzed the dynamic behavior and revealed interesting morphological changes with increasing driving force and/or droplet volume. Reference~\cite{Podgorski2001Corners} employs small-contact-angle silicon oil droplets on an inclined fluoro-polymer coated substrate and describes a sequence of transitions that droplets of fixed volume undergo with increasing substrate inclination (see Fig.~1 of Ref.~\cite{Podgorski2001Corners}): At small inclinations the droplets slide with constant speed and a constant oval-like shape of their base with a small front-back asymmetry (while at zero inclination they are shallow spherical caps). With increasing inclination angle the asymmetry between front and back increases, as does the droplet velocity. The front still resembles a circular arc but the back becomes increasingly pointed and develops a cusp at a critical capillary number (or non-dimensional droplet speed). Beyond the corresponding substrate inclination an instability occurs, where the sliding drop emits smaller satellite droplets at its back. This effect is also denoted as \textsl{pearling} \cite{Podgorski2001Corners}, a notion we also adopt here. In parallel, the shape of the unstable droplet develops an elongated protrusion at its back. The emitted droplets can be of identical size or follow a rich variety of dynamical periodic patterns. In this context, Ref.~\cite{Podgorski2001Corners} mentions a cascade of bifurcations that involve frequency divisions. It is found that the dependence of the sliding speed on inclination is nearly linear with a change of the corresponding slope at the onset of pearling. Power law dependencies on volume are also studied. Similarly, Ref.~\cite{Kim2002Sliding} derives and experimentally verifies for small droplet speeds (assumption that the droplet nearly remains a spherical cap) a power law that also includes dependencies on other experimental parameters. The limit of nearly non-wetting drops rolling down an incline is investigated in Refs.~\cite{MaPo1999pf,RiQu1999el}.

It is important to mention that the experiments in Ref.~\cite{Podgorski2001Corners} are actually not performed with individual sliding drops but with a continuous sequence of identical drops that are placed at a defined constant time interval at the top of the inclined plate. This does not affect the behavior below the threshold for the pearling instability. However, it implies that beyond the threshold one has pearling events where satellite drops split off the main drop and also coalescence events where main drops absorb satellite droplets left behind by the preceding main drop. This allows one to consider the pearling process together with coalescence events as a periodic process either in the fixed laboratory frame or in a frame moving with the main drops. This observation will below be used where we employ a long-wave model to analyze the morphological changes of the droplet shape and also the pearling-coalescence cycle. Also note that a similar change in droplet morphology (up to the appearance of a cusp) had earlier been observed for Marangoni-contracted water drops sliding down inclined hydrophilic substrates \cite{HuMa1991l}.

To model and simulate the sliding droplets, different approaches might be followed: One may employ the fully nonlinear Navier-Stokes equations describing the evolution of the full three-dimensional velocity field \cite{Wind-Willassen2014}. In most cases, describing slowly sliding drops, the linearized small-Reynolds-number form -- the Stokes equation -- may be used \cite{Dimi2007jfm}. Both approaches have to be complemented with appropriate boundary conditions at the solid-liquid and the liquid-air interface. Alternatively, one may combine the bulk equations with an additional phase-field dynamics, thereby modeling the liquid-air interface as diffuse \cite{DiSp2007pre,BoBB2008pre}. Usually, these approaches involve a rather high computational cost, in particular, if three-dimensional drops are considered. Other methods include dissipative particle dynamics \cite{JLRS1999fd} and lattice-Boltzmann methods \cite{HKRT2007epje}.

Alternatively, one may employ an asymptotic model derived from the Navier-Stokes and continuity equations and boundary conditions for free surface flows for which all typical length-scales parallel to the substrate are large as compared to the film height, i.e., for droplets with small contact angles and surface slopes. This small-gradient or long-wave approximation results for three-dimensional drops in a highly nonlinear evolution equation for the film thickness profile $h(\mathbf{r},t)$ as a function of the two coordinates $\mathbf{r} =(x,y)$ along the planar substrate \cite{Oron1997Long,Thie2007}. In the present case the resulting equation accounts for capillarity through a Laplace pressure and for wettability through a Derjaguin (or disjoining) pressure \cite{Genn1985rmp,Israelachvili2011}, that combines two antagonistic power laws \cite{pismen2001nonlocal,Thie2010jpcm}. When adding the lateral driving force due to gravity, they serve to study droplets sliding down an inclined homogeneous substrate.
Such long-wave models with various forms of Derjaguin pressure were already used to thoroughly analyze morphological transitions of stationary sliding two-dimensional drops (2d), i.e., with one-dimensional profiles $h(x,t)$ \cite{TVNB2001pre,Thiele2002Sliding}; and for a few time simulations of three-dimensional droplets (3d) that show the pearling instability \cite{BeNe2001prl,Thiele2002Sliding,Schwartz2005Shapes} or states close to pearling \cite{Pesc2015jcp}. Such models are also employed to study the combined spreading and sliding of perfectly wetting drops \cite{Schw1989pfa,Oron1997Long}.

Note that we neglect any roughness of the substrate beneath the sliding droplets. Therefore, any contact angle hysteresis known from experiments is excluded from the dynamics that we observe, and there is no finite onset inclination angle at which the droplets start to move. The related depinning behavior is observed with similar long-wave models where modulations of the Derjaguin pressure represent regular arrays of wettability defects in Refs.~\cite{ThKn2006njp} (2d) and \cite{BeHT2009el,BKHT2011pre} (3d). Long-wave equations are also employed to study droplet spreading over random topographical substrates \cite{SaKP2010prl}.
The influence of regular and random substrate modulations on the onset of droplet sliding is also studied with lattice-Boltzmann methods in Refs.~\cite{SBAV2014pre} and \cite{HKRT2007epje}, respectively. Here, we always assume an ideally smooth substrate as can, e.g., be realized through atomically flat single crystal surfaces or through liquid-infused substrates.

The present work employs such an asymptotic long-wave model to investigate the morphological changes of fully three-dimensional sliding drops as observed in Ref.~\cite{Podgorski2001Corners}. This is done employing \textsc{pde2path}, a continuation and bifurcation package for elliptic systems of partial differential equations (PDEs) \cite{P2P2014,UeWe2014sjads}. In particular, we study the behavior of sliding droplets at several fixed drop volumes as the 
 inclination angle of the substrate is changed that represents the strength of the driving force. Our analysis shows that a family of stationary sliding droplet solutions exists that undergoes steady and oscillatory bifurcations. These bifurcations are related to morphology changes as, e.g., the development of protrusions at the droplet back (related to a pair of saddle-node bifurcations) and the emergence of satellite droplets in a pearling instability that technically is a global (homoclinic) bifurcation. Beside the bifurcation diagram of stationary droplets we obtain their linear stability and the relevant eigenmodes. We obtain predictions on existence ranges and instabilities that are tested via direct numerical simulations that are also used to investigate the branch of time-periodic behavior (corresponding to pearling-coalescence cycles) which emerges at the global instability, the related hysteresis in behavior, and a period-doubling cascade that occurs when further increasing the inclination. The non trivial oscillatory behavior close to a Hopf bifurcation of drops with finite-length tails is also studied. Finally, it is shown that the main features of the bifurcation diagram are universal as they follow scaling laws over several decades of the droplet size. The quantitative changes are condensed in the form of scaling laws. Finally, we consider the total dissipation of the different droplet types and analyze where dissipation is localized within the sliding droplets.

The paper is structured as follows. In Sec.~\ref{sec:mod} we present our model and discuss the numerical approaches of time-stepping and path-continuation. The sub-sections of Sec.~\ref{sec:results} present our main results, first, in the form of the overall bifurcation diagram and, then, detailing the behavior on the various found sub branches of stationary sliding drops. Sec.~IV analyses the dependence of main features on the droplet volume while Sec.~V discusses the velocity field and dissipation within the drops. The work concludes with a discussion in Sec.~\ref{sec:concl}.

\section{Modelling and Numerical Approach}
\label{sec:mod}
\subsection{Governing equation}
\label{sec:model}
We describe the dynamics of sliding drops using an asymptotic long-wave model, i.e., an evolution equation for the film thickness profile. The model accounts for capillarity through a Laplace pressure, wettability through a Derjaguin (or disjoining) pressure and the down-hill component of gravitation as lateral driving force \cite{Oron1997Long,Thiele2002Sliding,Schwartz2005Shapes}. It is derived from the Navier-Stokes equations using a long-wave approximation \cite{Oron1997Long,Thie2007} employing a no-slip boundary condition at the smooth solid substrate and stress-free boundary conditions at the liquid-air interface, i.e., the free surface of the film.
This approach is valid in the case of thin liquid layers that have small surface slopes, i.e., for liquid films and drops that are described by a height profile $h(x,y,t)$ which only exhibits small gradients in the lateral directions. In other words, the surface deflections are small (or the entire film is thin) relative to the lateral dimensions of height modulations.\par 
For a general lateral driving force $\boldsymbol{\chi}(h)$, the non-dimensional evolution equation for the film thickness profile reads:
\begin{align}
\partial_t h(x,y,t) =&\ -\nabla\cdot\left[Q(h)\nabla\left[-\Delta h(x,y,t) + \Pi(h) \right] + \boldsymbol{\chi}(h) \right] \label{TFEmodel}
\end{align}
where $Q(h) = h^3$ is the mobility function and $\Pi(h)= - df(h)/dh$ is the Derjaguin pressure modeling wettability, i.e., the effective interaction between the free surface of the film and the substrate \cite{Genn1985rmp,StVe2009jpm}. The term $\Delta h$ represents the Laplace pressure due to curvature. We employ the wetting potential $f(h) = \frac{1}{5h^5}  - \frac{1}{2h^2}$ that corresponds to a precursor film model, i.e., a macroscopic drop coexists at equilibrium with an thin adsorption layer of thickness $h_\mathrm{p} = 1$ determined by $\Pi(h_\mathrm{p})=0$ \cite{pismen2001nonlocal,Thie2010jpcm}. Note that there are many other possible choices for the wetting potential and resulting Derjaguin pressure. Another common ansatz employs, e.g., exponential functions \cite{BeNe2001prl,Thiele2002Sliding} or a combination of a power law and an exponential \cite{Sharma1993,ThVN2001prl}, however the resulting overall shape of $\Pi(h)$ is rather similar. Instead of using a Derjaguin pressure that ensures the existence of a wetting layer, other approaches are possible, such as a slip-model that allows for a finite slip velocity of the liquid at the liquid-solid interface \cite{Hock1977jfm}. Although there exist quantitative differences to the model used here, time simulations show qualitatively similar results for individual sliding drops (see, e.g., Ref.~\cite{Pesc2015jcp}). Furthermore one can show that asymptotically these models are equivalent in describing moving contact lines \cite{SaKa2011el}. However, for the current work, a slip model is less suitable because additional assumptions are needed to treat topological changes of drop solutions with compact support. Such changes frequently occur during the studied pearling-coalescence cycles.

In the non dimensional form of the thin film equation in Eq.~(\ref{TFEmodel}), all material parameters, such as viscosity, surface tension, and Hamaker constant, are already absorbed into the scaling of height, the coordinates, and time, resulting in the remaining dimensionless parameters (as described in note \footnote{Given the dimensional form of the disjoining pressure \unexpanded{$\widetilde{\Pi}(\tilde h) = -A/\tilde h^3 + B/\tilde h^6}$ (where $A$ is a Hamaker constant and $B$ is the interaction strength of the short-range contribution) the film height is scaled as $\tilde h = h_\mathrm{eq} h$ with $h_\mathrm{eq} = (B/A)^{1/3}$ defining the precursor layer height and length scale in the $z$ direction. This implies an equilibrium contact angle $\theta_\mathrm{eq} = \sqrt{3A/(5\gamma h_\mathrm{eq}^2)}$. After introducing the $x$ and $y$ length scale $l_0 = \sqrt{3/5}\,h_\mathrm{eq}/\theta_\mathrm{eq}$, the time scale $t_0 = (9\eta h_\mathrm{eq})/(25\gamma\theta_\mathrm{eq})$, and the scales for the driving force $\tilde{G}_0 = A/(\rho h_\mathrm{eq}^4)G_0$ and $\alpha_0 = h_\mathrm{eq}/l_0$ one obtains the non-dimensional evolution equation (\ref{TFEmodel}). The resulting velocity scale is $u_0 = l_0/(3t_0)$.\\ Starting with the non-dimensional form used here, one may retrieve dimensional quantities by introducing specific interaction constants $A$ and $B$ as well as surface tension $\gamma$ and viscosity $\eta$. With this, length and time scales as well as the equilibrium contact angle $\theta_\mathrm{eq}$ and wetting layer thickness $h_\mathrm{eq}$ are determined. Alternatively, one may give $\theta_\mathrm{eq}$ and $h_\mathrm{eq}$ instead of $A$ and $B$.}). 
For details of the used scaling, we refer to Refs.~\cite{Thiele2002Sliding,ThKn2004pd,Thie2010jpcm,tseluiko2014collapsed}. Note that the scaling is chosen in such a way that each of the most important control parameters only enters the equation through one dimensionless parameter but is not contained in the scales. Here, these is 
the inclination of the substrate and the volume of the drop.

In the case of an inclined substrate, the influence of gravity is incorporated into the driving force $\boldsymbol{\chi}(h)$ through the 
scaled long-wave inclination angle $\tilde \alpha$:
\begin{align} \label{drivingforce}
\boldsymbol{\chi}(h) =&\ Q(h) G_0 \ (\tilde \alpha, 0)^T.
\end{align}
Here, $G_0 = 10^{-3}$ presents a dimensionless gravity parameter and the plane is tilted in the $x$ direction; i.e., there is no driving force in the $y$ direction. For simplicity, we always use $\alpha = G_0 \tilde \alpha$ as an effective inclination.

We note, that Eq.~(\ref{TFEmodel}) with driving (\ref{drivingforce}) may be expressed in the general variational form of a gradient dynamics on a free energy functional $\mathcal{F}[h]$. Namely,
\begin{equation}
 \partial_t h = \nabla\cdot\left[ Q(h) \nabla \frac{\delta \mathcal{F}[h] }{\delta h}\right]
\end{equation}
with
\begin{equation}
 \mathcal{F}[h] 
 = \int_\Omega \left[ \frac{1}{2}(\nabla h)^2 + f(h) \right] \, dx\,dy - \int_\Omega  h \alpha x \, dx\,dy\, .
\label{eq:ener}
\end{equation}
The final term in $\mathcal{F}[h]$ corresponds to the lateral part of the potential energy. Note, that (\ref{eq:ener}) has no lower bound, as the global minimum is only reached if all the liquid volume approaches $x \rightarrow \infty$. Therefore, we have the interesting case of a system that remains out of equilibrium  forever, but is still described by a gradient dynamics that evolves towards an energy minimum (it can be shown that $\mathcal{F}[h]$  is a Lyapunov functional). This dramatically distinguishes the case treated here from the relaxational dynamics discussed in Refs.~\cite{Mitl1993jcis,Thie2010jpcm} where the system moves towards equilibrium and approaches it.\par 

To investigate droplets that slide with a constant shape and velocity $U$, i.e., stationary solutions of Eq.~\eqref{TFEmodel}, the equation is transformed into a comoving frame $\tilde x=x-Ut$ and the time derivative in the comoving frame is set to zero. Practically, this replaces the time derivative in Eq.~\eqref{TFEmodel}  by an advection term $- (U, 0)^T\cdot\nabla h$. Dropping the tilde we obtain the equation
\begin{align}
0 =  -\nabla\cdot\left[Q(h)\nabla\left[\Delta h(x,y,t) + \Pi(h) \right] + \boldsymbol{\chi}(h) - (U, 0)^T h \right]
\label{TFEcomoving}
\end{align}
that determines the stationary states $h=h_0(x,y)$. To determine their stability, we add a small perturbation $h_1(x,y)$, i.e., use the ansatz $h(x,y,t)= h_0(x,y)+h_1(x,y)\exp(\lambda t)$. Introducing the ansatz into Eq.~(\ref{TFEmodel}) and linearizing in $h_1$ gives (in the comoving frame) 
the linear eigenvalue problem
\begin{align}
 0 =&\ \left(\mathcal{L}[h_0] - \lambda \right)h_1\,,
 \end{align}
 where $\lambda$ and $h_1$ represent the eigenvalue and eigenfunction, respectively.
 The linear operator $\mathcal{L}[h_0]$ is defined by
 \begin{align*}
 \mathcal{L}[h_0] h_1 =&\ -\nabla\cdot\left[h_1\,3h_0^2 \nabla\left[\Delta h_0 + \Pi(h_0) \right] + ( 3h_0^2 \alpha -U,0)^T h_1 + h_0^3 \nabla\left[\Delta h_1 + \Pi'(h_0) h_1 \right] \right]\,.
\end{align*}
The presented model has a variety of different solution types. A homogeneous film of any height is always a (trivial) solution. However, for the employed Derjaguin pressure flat films are unstable in most cases (linearly unstable for $h>2^{1/3}$). Also, effectively one-dimensional solutions are possible, such as transversally invariant ridges that slide down the incline. Their cross sections in the $x$ direction are one-dimensional sliding drops, i.e., $h=h(x,t)$. Another option are rivulets that represent free surface channels guiding the liquid downslope. They have a cross section in the $y$direction independent of $x$, i.e., $h=h(y,t)$.
Here, we mainly analyze truly three-dimensional droplets, i.e., $h=h(x,y,t)$, using both, continuation techniques and direct numerical simulations, which are briefly described next.

\subsection{Numerical approach}
\subsubsection{Time stepping}
\label{sec:dns}
Direct numerical simulations of the model \eqref{TFEmodel} are performed using a finite-element method. The method is implemented through the open source framework \textsc{dune-pdelab} \cite{bastian2010generic,bastian2008genericI,bastian2008genericII}. We discretize the simulation domain $\Omega = [0,L_x]\times [0,L_y] = [0,200]\times [0,100]$ into an equidistant mesh of $N_x\times N_y = 128 \times 64$ quadratic elements with linear (Q1) ansatz and test functions. The time integration is conducted with an implicit second-order Runge-Kutta scheme \cite{alexander1977diagonally}. The resulting nonlinear problems are solved with a Newton method, using for the linear problems a biconjugate-gradient-stabilized method (BiCGStab) with a symmetric successive overrelaxation (SSOR) as preconditioner.  
To implement Eq.~\eqref{TFEmodel} using this ansatz, we split the model into two equations second order in space:
\begin{align}
 \partial_t h =& - \nabla \cdot \left[Q(h) \nabla [w + \Pi(h)] + \chi(h)\right] \label{TFEdns}\,, \\
 w =& \Delta h\,.
\end{align}
In a weak formulation for general test functions $\phi_h$ and $\phi_w$ the equations read
\begin{align}
0 =& - \int_\Omega \partial_t h \phi_h \, dx\,dy  +\int_\Omega  Q(h) \nabla (w+\Pi(h)) \cdot \nabla \phi_h + \chi(h) \cdot \nabla \phi_h \, dx\,dy, \\
 0 =& \int_\Omega w \phi_w + \nabla h \cdot \nabla \phi_w  \, dx\,dy.
\end{align}
The boundary integrals occurring due to the performed partial integrations vanish due to the used periodic boundary conditions.
As initial conditions we use a parabolic cap of prescribed volume $V$ above the precursor layer. When the time-dependent behavior of unstable solutions is studied, solutions obtained by continuation (Sec.~\ref{sec:path_cont}) are used as initial conditions.\par
To obtain the velocity  of the sliding droplets, we track the position of maximal height over time and differentiate numerically using fourth order finite differences. To better represent the results in the forthcoming plots of height profiles, the $x$ coordinate of the solutions is shifted such that the maximum is always at the same position. This corresponds to a transformation into the frame moving with the numerically determined droplet velocity.

\subsubsection{Path-continuation}
\label{sec:path_cont}
To analyze the model \eqref{TFEmodel} with continuation methods \cite{Kuznetsov2010,DWCD2014ccp}, i.e., to directly track solutions in parameter space, we make use of the numerical pseudo-arclength continuation package \textsc{pde2path} \cite{UeWR2014nmma,P2P2014} which is based on the finite-element methods of \textsc{matlab}'s \textsc{pdetoolbox}. \par 
Basically, path continuation determines steady states of an ordinary differential equation (ODE) combining prediction steps where a known solution is advanced in parameter space via a tangent predictor and correction steps where refined Newton procedures are employed to converge to the new solution at a new value of the primary continuation parameter. PDEs including boundary conditions and side conditions as, e.g., volume conservation can always be approximated by an ODE system of large dimension. The primary continuation parameter is in our case, e.g., the inclination $\alpha$ of the substrate in Eq.~\eqref{TFEmodel}. Naturally the continuation of sliding drops requires the use of the system in the comoving frame as given by Eq.~\eqref{TFEcomoving}, implying that beside the primary continuation parameter also the velocity of the frame has to be determined. In this way one may start at an analytically or numerically given solution, continue it  in parameter space and obtain a broad range of solution families including their bifurcations and accompanying changes in morphology \cite{DWCD2014ccp,Kuznetsov2010}. However, it depends on the continuation package whether only time-independent solutions can be continued or whether this is also possible for time-periodic states. As the latter is often not the case, direct numerical simulations are needed to complete the bifurcation diagram of a system.

In time simulations of \eqref{TFEmodel}, the volume is automatically conserved because the equation has the form of a continuity equation, $\partial_t h = -\nabla \cdot \mathbf{j}$. To ensure conservation of volume while following stationary solutions in parameter space via continuation, an additional integral constraint on the solutions is needed (see below).\par
To increase computational efficiency, we exploit the symmetry of the drops perpendicular to the inclination direction and only compute one half of the physical domain $\Omega = [0,L_x]\times [0,L_y] = [0,200]\times [0,100]$, i.e., $\Omega_\mathrm{num} = [0,L_x]\times [0,L_y/2]$. $\Omega_\mathrm{num}$ is discretized on a grid with $N_x\times N_y = 800 \times 200$ mesh points. As a result Neumann conditions are imposed on the boundaries in the $y$ direction, whereas periodic boundary conditions are imposed in the $x$ direction; i.e., we investigate periodic arrays of sliding drops. \par
Further, one needs an additional phase condition to break the translational invariance in the $x$ direction. The condition prevents the continuation algorithm to trivially follow solutions along the translational degree of freedom. Volume is fixed during continuation by adding a volume condition. The two conditions read
\begin{equation}
 \int_{\Omega} h \, \partial_x h_\mathrm{old}\, dx\,dy = 0 \qquad \text{and} \qquad \int_{\Omega} (h - h_\mathrm{p})\, dx\,dy - V = 0,
\end{equation}
respectively. Here $h_\mathrm{old}$ denotes the solution obtained in the previous continuation step. \par
Taking a single static drop with $\alpha = U = 0$ as starting solution, we conduct an pseudo-arclength continuation for increasing inclination angles $\alpha$. The corresponding velocity $U$ of the comoving frame is used as an additional free parameter, as the drops slide with varying velocities that depend on the inclination and are automatically adapted to the corresponding $\alpha$ during continuation.

  \begin{figure}[htbp]
  \center
     \includegraphics[width=1.\textwidth]{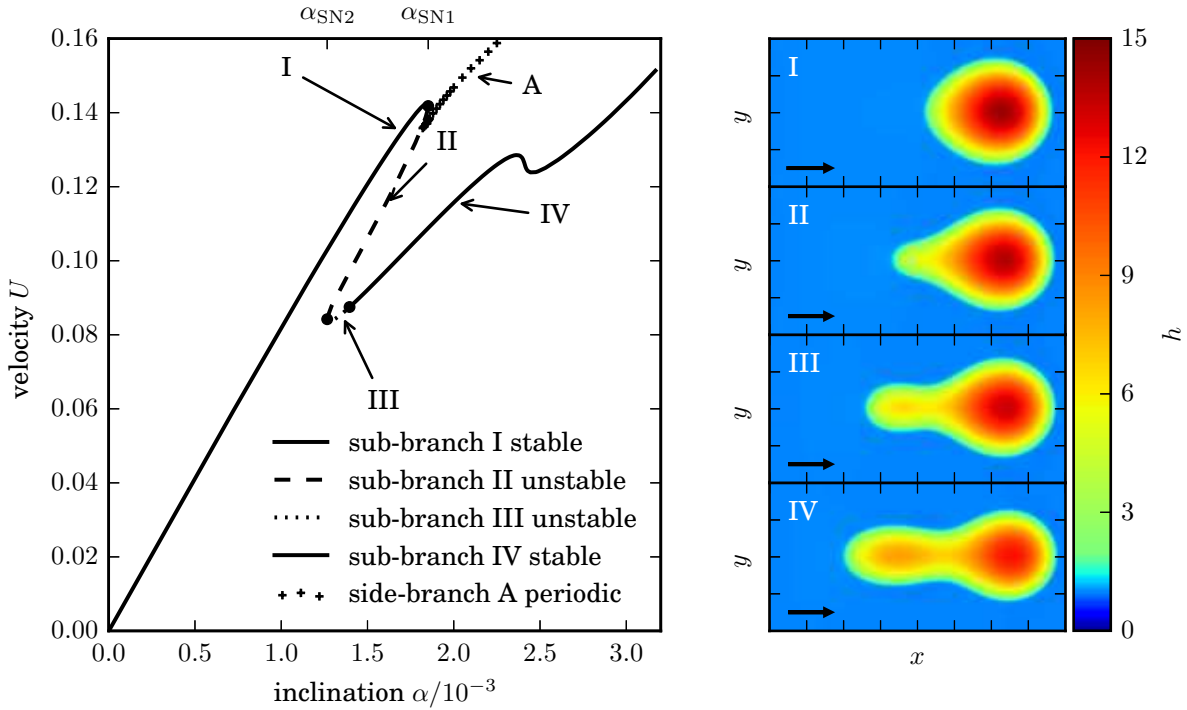}
  \caption{The left panel shows the bifurcation diagram in terms of the dependence of the velocity $U$ of stationary sliding drops (solid and dashed lines) on inclination angle $\alpha$ at fixed volume $V = 3.0\times10^4$. The right panels show exemplary stationary drop profiles on a domain $\Omega = 200\times100$ from the various sub-branches as indicated by roman numbers in the left panel. The arrows indicate the sliding direction. Side-branch A consists of time-periodic solutions obtained by direct numerical integration and is discussed below in Sec.~\ref{sec:branchA}, while snapshots of these solutions are shown in Fig.~\ref{fig:fig2_pearling}. In the left panel, the time-averaged velocity of these solutions is shown. An animation showing the changing shape of the stationary droplets as one travels along the bifurcation curve is available as movie 01 in the Supplementary Material.}
  \label{fig:fig1_bifurcation}
  \end{figure}  

\section{Bifurcation Diagram}
\label{sec:results}
At first, we give an overview of changes in the overall droplet behavior as the inclination angle is increased. The individual phenomena are further analyzed in the subsequent sections. \par
The bifurcation diagram shown in Fig.~\ref{fig:fig1_bifurcation} summarizes all stationary droplet states obtained by continuation. It presents the dependency of the velocity $U$ on inclination angle $\alpha$ for droplets of fixed volume $V=3\times10^4$. The main branch displayed by solid and dashed lines represents the different stationary drops. The general form of the branch with its prominent folds is universal over a wide range of drop volumes $V$ (discussed in Section \ref{sec:powerlaws} below). Depending on the shape of the bifurcation curve and the linear stability of the individual solutions, we divide the branch into sub-branches consisting of drop profiles that show common properties. Note that at this point, we discuss only the morphology and overall behavior of the different droplet types. For a further analysis of the velocity field and dissipation within the droplets see Sec.~\ref{sec:dissipation}.

From the main branch of stationary states, a side branch of \textsl{pearling} solutions bifurcates, that represents the emission of small satellite drops from the larger sliding drops. Due to the periodic boundary conditions in the direction of droplet motion, these solutions are time-periodic in the comoving frame, which corresponds to the experimental setup in Ref.~\cite{Podgorski2001Corners} where a continuous sequence of identical drops slides down an incline. The distance between the drops corresponds to the periodicity of the domain considered here.\par
Starting with a droplet at rest for a vanishing inclination ($\alpha=0$), the first part of the curve $U(\alpha)$ (denoted in the following as \textit{sub-branch I}) exhibits a linear increase of velocity with increasing inclination angle. The overall shape of the drop profile does not change significantly (cf.~Fig.~\ref{fig:fig1_bifurcation}, snapshot I; for details see Sec.~\ref{sec:branchI}).
For a specific value of $\alpha = \alpha_\mathrm{SN1}$, which depends on the drop volume, a saddle-node bifurcation (or fold) occurs where the stable sub-branch~I connects with a linearly unstable sub-branch~II. The latter continues towards smaller $\alpha$; i.e.,  it coexists with sub-branch~I for a range of inclination angles and always exhibits smaller velocities. The shape of the droplets changes significantly along this sub-branch, as a protrusion  is formed at the rear of the drop (cf.~Fig.~\ref{fig:fig1_bifurcation}, snapshot II) that is the mesoscopic equivalent of cusps observed for macroscopic drops. This is further discussed in Sec.~\ref{sec:branchII_III}.
Near the saddle-node bifurcation, on sub-branch~II also a global bifurcation occurs, that is responsible for the pearling instability and the emission of small satellite droplets. The resulting non-stationary solutions are time-periodic and show a repeated breakup of the main drop into a large main droplet and a small satellite droplet and the subsequent coalescence of the two. This type of behavior is determined by direct numerical simulations and is denoted as \textit{side branch~A}. It is further analyzed in Sec.~\ref{sec:branchA}.\par
Following sub-branch~II with decreasing $\alpha$, at a lower inclination threshold $\alpha = \alpha_\mathrm{SN2}$, a second saddle-node bifurcation occurs. There, a second real eigenvalue becomes positive. Therefore beyond the fold, ``sub-branch~III'' consists of doubly unstable drops and continues towards larger $\alpha$. 
The two real unstable eigenvalues approach each other, become identical already slightly after the saddle-node bifurcation. Then they transform into a complex conjugate pair. Beyond this point, time simulations  that start with such unstable steady drops on sub-branch~III  show \textsl{breathing}-like oscillatory destabilization and then approach a drop on the stable upper sub-branch~I. Increasing the inclination angle further, the pair of complex conjugate eigenvalues crosses the imaginary axis, i.e., a Hopf bifurcation occurs. At this bifurcation, we expect a branch of time-periodic solutions to emerge, e.g., periodically oscillating droplets. However, the employed continuation package is not yet able to follow such a branch of time-periodic solutions. As our direct time simulations in this parameter range did not show such solutions, we conjecture that the branch of time-periodic solutions is a short (small $\alpha$ range) unstable branch that starts subcritically at the found Hopf bifurcation and ends in a homoclinic bifurcation (as sometimes seen in related systems; see~e.g.~Ref.~\cite{LRTT2016pf}). Note, that this hypothetical short branch should not be confused with the aforementioned side-branch~A of periodic pearling and coalescence that emerges at a different homoclinic bifurcation. Beyond the Hopf bifurcation, we are on sub-branch~IV that is formed by linearly stable droplets. The protrusion at the back of the droplets becomes even more significant in this region as an elongated tail is formed whose length increases with increasing inclination angle (cf.~Fig.~\ref{fig:fig1_bifurcation} snapshots III/IV). Note that depending on $\alpha$ this non-trivial droplet solutions coexists with either the stable drops of sub-branch~I or with the stable time-periodic behavior on side-branch~A. Increasing $\alpha$ even further, on branch~IV the droplet elongates until it forms a rivulet solution in the $x$ direction (see Sec.~\ref{sec:branchIV} for details). In the next sections we further analyze the phenomena on the individual sub-branches.

\subsection{Stable stationary sliding drops: Sub-branch~I}
\label{sec:branchI}
First we give some more details on sub-branch~I, i.e., the part of the bifurcation curve in Fig.~\ref{fig:fig1_bifurcation}, that emerges from the origin. Due to the absence of substrate inhomogeneities or other pinning mechanisms, a drop at rest begins to slide for arbitrarily small finite values of $\alpha$ (cf.~Refs.~\cite{ThKn2003pf,ThKn2004pd} and the present case in comparison to Refs.~\cite{ThKn2006njp,BKHT2011pre}). Inclining the substrate further, the drop velocity increases linearly while the shape of the drop does not vary significantly. The radially symmetric spherical cap shape at $\alpha=0$ becomes slightly oval as the droplet prolongs a bit in the direction of the incline. Further, one can observe that the $x\to-x$ symmetry is broken as the trailing back of the drop becomes a bit more pointed than the advancing front (cf.~Fig.~\ref{fig:fig1_bifurcation}, snapshot I). The value of the slope of the linear dependence $U(\alpha)$ depends on the volume of the drop, and can be brought into the form of a simple power law. This is quantitatively analyzed below in Sec.~\ref{sec:powerlaws}. The final part of sub-branch I at $\alpha \gtrsim 1.5\times 10^{-3}$ deviates from the linear behavior towards lower velocities until it becomes vertical where it ends in the first saddle node at $\alpha_\mathrm{SN1} \approx 1.85\times 10^{-3}$. In this final part of the sub-branch the shape asymmetry of the droplets gets stronger.  The dependencies of the critical angle $\alpha_\mathrm{SN1}$ and the velocity $U_\mathrm{SN1}$ at the bifurcation on droplet volume can also be brought into the form of power laws as discussed below in section \ref{sec:powerlaws}.

\subsection{Time-periodic behavior - Side branch~A}
\label{sec:branchA}

\begin{figure}[htbp]
  \center
     \includegraphics[width=1.\textwidth]{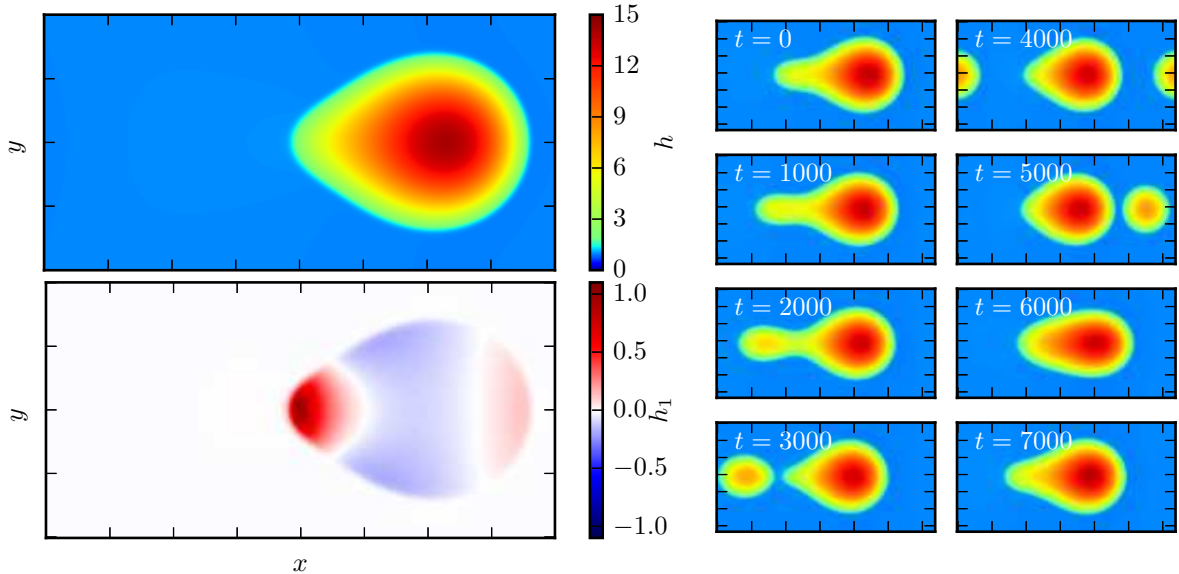}
  \caption{Stationary sliding droplet (top left panel) and eigenfunction of marginally stable perturbation mode (bottom left panel) at the first saddle-node bifurcation ($\alpha=\alpha_\mathrm{SN1}\approx 1.85 \times 10^{-3}$) in the bifurcation diagram, Fig.~\ref{fig:fig1_bifurcation}. The panels on the right show snapshots of the time evolution for an inclination angle $\alpha = 1.96\times 10^{-3}$ slightly above the saddle-node bifurcation. A video of this simulation is available as movie 02 in the Supplementary Material.}
  \label{fig:fig2_pearling}
\end{figure}

As shown in the bifurcation diagram in Fig.~\ref{fig:fig1_bifurcation}, the sub-branch of stable sliding droplets of weak deformation from the equilibrium spherical cap profiles ends in a saddle-node bifurcation at $\alpha=\alpha_\mathrm{SN1}$. At this bifurcation a single real eigenvalue becomes positive. The corresponding eigenfunction and the stationary droplet profile close to the bifurcation are shown in Fig.~\ref{fig:fig2_pearling}. 

For inclination angles larger than $\alpha_\mathrm{SN1}$, there exist no stationary sliding droplets with a simple spherical-cap-like shape. Instead, one may find a periodic (or more involved) spatio-temporal dynamics. The continuation package we employ is tailored to follow steady states of elliptical problems and can therefore not be employed in the analysis of such a spatio-temporal dynamics. To determine what type of time-varying behavior occurs for $\alpha>\alpha_\mathrm{SN1}$, we use direct numerical simulations (DNS) of Eq.~(\ref{TFEmodel}).

As initial condition for the DNS, we use the stable solutions on sub-branch~I below but close to $\alpha_\mathrm{SN1}$, and increase the inclination angle to an $\alpha > \alpha_\mathrm{SN1}$ above the critical one. It turns out that the unstable eigenfunction shown in the bottom left panel of Fig.~\ref{fig:fig2_pearling} also corresponds to the dominant mode of evolution. It results in a \textsl{pearling} behavior: The large main drop ejects from its pointed backside satellite droplets of considerably smaller volume than its own. 
As the smaller drop moves more slowly than the larger one, the distance between them continuously increases. 
Due to the periodic boundary conditions in the $x$ direction, after a certain time period $T_\mathrm{per}$ the small drop is absorbed again by the main drop. The resulting merged drop is then again unstable and emits another satellite droplet; i.e., the behavior corresponds to a limit cycle solution. At first view, this periodic behavior might seem artificial as it is induced by the periodic boundary conditions. However, these conditions are appropriate also for the real experimental system \cite{Podgorski2001Corners} where one periodically places identical drops at the top end of an inclined plate. A constant frequency is obtained by continuously dripping liquid from a nozzle. This frequency $\omega$, in combination with the mean sliding velocity of the drop $\bar v$, directly translates to a spatial periodicity $\omega^{-1} \bar v$ corresponding to the size $L_x$ of our spatial domain with periodic boundary conditions. The experimental dripping frequency therefore scales as $ L_x^{-1}$ for large domain sizes. Then, the duration of the droplets interaction becomes negligible compared to the time interval in which the two individual drops freely slide without interaction.

\begin{figure}[htbp]
  \center
     \includegraphics[width=0.7\textwidth]{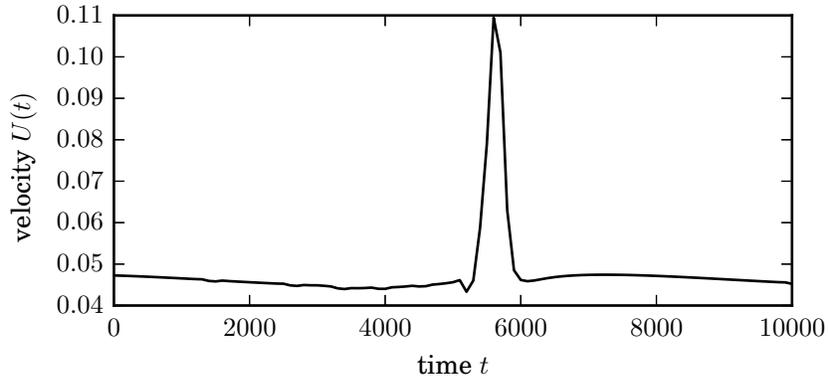}
  \caption{Velocity of the maximum of the larger drop during the pearling process with time period $T \approx 7780$. The time $t$ corresponds to the time displayed in the snapshots of Fig.~\ref{fig:fig2_pearling}. The large peak in the middle occurs during the coalescence process, where the larger drop quickly invades the smaller one. Shortly after this, at $t \approx 7000$, the drop has gained a maximal compact shape, where it slides fastest. During the following elongation and formation of the tail, it slows down until the pearling occurs at $t \approx 3000$. The two resulting drops then relax again to a more compact shape, resulting in a increase of sliding velocity around $t \approx 4000$.}
  \label{fig3_pearling_velocity}
\end{figure}

To be able to include the branch of these time-periodic solutions into the same bifurcation diagram in (Fig.~\ref{fig:fig1_bifurcation}) that shows the stationary sliding drops, we compute the velocity of the larger drop averaged over ten periods $T_\mathrm{per}$, for which we track the position of maximal drop height over time and differentiate it numerically. 
Figure~\ref{fig3_pearling_velocity} shows that the droplet velocity greatly varies over the course of one time-period because the point of maximal height moves faster for the fused drop than for the split drops. The position of the maximum moves even faster during the coalescence process. The temporal average of this velocity also depends on the spatial domain size that is considered, because for a  larger domain size, after the pearling it takes longer for the larger drop to catch up with the satellite drop. As for the  separated drops the velocity is smaller than during the coalesced phase and a larger domain size results in a smaller average velocity.

\begin{figure}[htbp]
  \center
     \includegraphics[width=1.\textwidth]{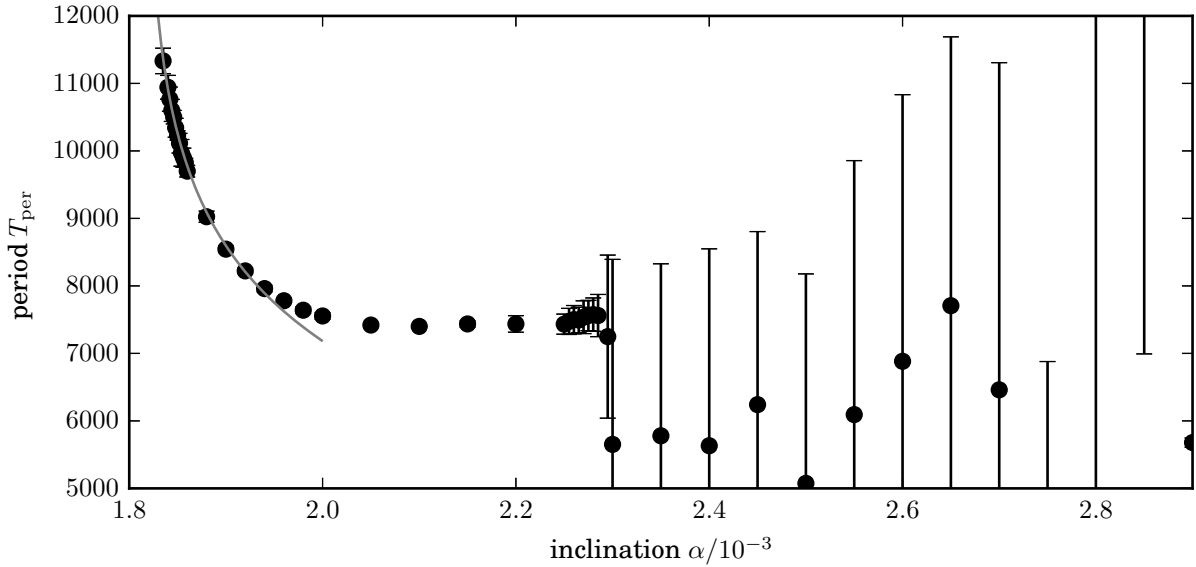}
  \caption{Time period of the pearling-coalescence cycle in dependence of the inclination angle $\alpha$. The period is averaged over 10 cycles and the resulting standard deviation is indicated as error bars. Towards the onset of pearling at lower $\alpha$, the period scales as $T_\mathrm{per} \sim -\mathrm{ln}(\alpha -\alpha_\mathrm{bif})$, as indicated by the gray fit line. After the first period doubling at $\alpha \approx 2.15 \times 10^{-3}$, the standard deviation increases significantly, indicated by the error bars. The remaining parameters are as in Fig.~\ref{fig:fig1_bifurcation}.}
  \label{fig:fig4_pearling_periods}
\end{figure}

Figure \ref{fig:fig4_pearling_periods} shows how the time-period $T_\mathrm{per}$ of the pearling-coalescence cycle depends on the inclination angle (for a fixed domain size). From the bifurcation point at $\alpha_\mathrm{bif}$ the period is quickly decreasing with increasing angle with a logarithmic scaling $T_\mathrm{per} \sim -\mathrm{ln}(\alpha -\alpha_\mathrm{bif})$, which is typical for an onset of a periodic cycle via a homoclinic bifurcation \cite{Strogatz1994}. This hypothesis is also supported by the fact that there exists some hysteresis, i.e., in a small interval of inclination angles $\alpha_\mathrm{bif}<\alpha<\alpha_\mathrm{SN1}$ both, the time-periodic solutions and the stationary drops of sub-branch I are stable (see also Fig.~\ref{fig:fig7_bifurcationzoom} below).
This zoom also clarifies that the time-periodic branch ends on sub-branch~II of unstable droplets as expected for a homoclinic bifurcation. 

For larger inclination angles $\alpha \gtrsim 2.1 \times 10^{-3}$, the time-period $T_\mathrm{per}$ increases again. Also the standard deviation starts to increase  (Fig.~\ref{fig:fig4_pearling_periods}). This indicates that the period of the subsequent pearling-coalescence cycles is not identical anymore, but begins to vary. In fact, from $\alpha \approx 2.15 \times 10^{-3}$ period doublings occur: First, the period begins to alternate between two different values but for larger $\alpha$ more periods occur.

\begin{figure}[htbp]
  \center
     \includegraphics[width=1.\textwidth]{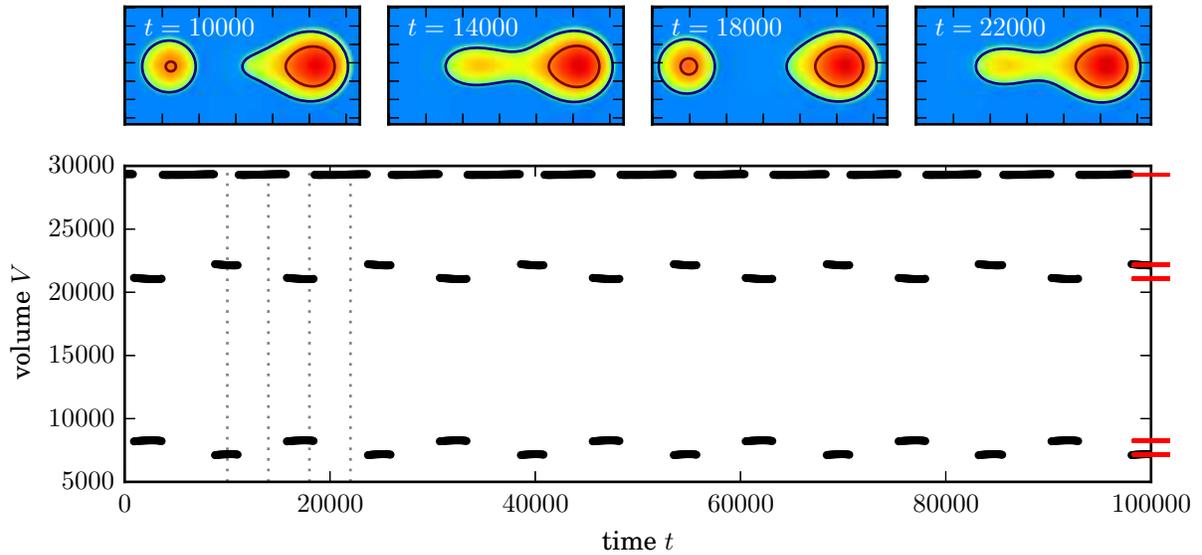}
  \caption{Bottom panel: time series of all droplet volumes present in a simulation of a pearling drop for $\alpha = 2.2\times 10^{-3}$. Fused states at large volume alternate with split-up states of smaller volumes, indicated by the repeated alternation between time intervals with one or two drop sizes. Two subsequent split-up states exhibit different drop volumes, while the fused states have all identical volumes, which is a sign of a period doubling. The dashed lines mark the particular times that correspond to the exemplary solution snapshots in the top panel. The small drop in the third snapshot (at $t=1.8\times10^4$) has a larger volume than the smaller drop at $t=1.0\times10^4$. The red bars at the right hand end of the bottom panel summarize all drop volumes that occur during the entire simulation.}
  \label{fig:fig5_pearling_volumes_timeseries}
\end{figure}

This period doubling can best be illustrated by studying the volumes of all drops after pearling, i.e., in the split-up state. To determine the volume of the drops, we identify connected areas in the simulation domain, that are elevated above a certain threshold height $h_\mathrm{threshold} = 1.05$ and integrate over this subdomain:
\begin{equation}
 V_\mathrm{drop} = \int_{\Omega_\mathrm{drop}} (h-h_\mathrm{threshold}) \,dx\,dy
\end{equation}
The threshold height is subtracted in the integration to ensure that drops of identical \textsl{real} volume, i.e., volume above the adsorption layer, but different base area give the same value of $V_\mathrm{drop}$. Although this results in a slight systematic underestimation of drop volumes, 
the effect is negligible as we do not quantitatively compare to other methods.

The bottom panel of Fig.~\ref{fig:fig5_pearling_volumes_timeseries} shows the time series of all drop volumes in a simulation for $\alpha = 2.2 \times 10^{-3}$. The repeated break up and coalescence is visible as alternating time intervals with one and two concurrent volumes. Two successive split-up states, e.g., at $t=1.0\times10^4$ and $t=1.8\times10^4$ exhibit different but alternating volumes. This alternation between two different split-up states indicates a period doubling of the pearling-coalescence cycle. Combining all occurring drop volumes of such a simulation (see the short red horizontal bars at the right edge of Fig. \ref{fig:fig5_pearling_volumes_timeseries}), one can classify whether the simulation shows a simple time-periodic state, a period-doubled state or states with more complicated time dependencies.

\begin{figure}[htbp]
  \center
     \includegraphics[width=1.\textwidth]{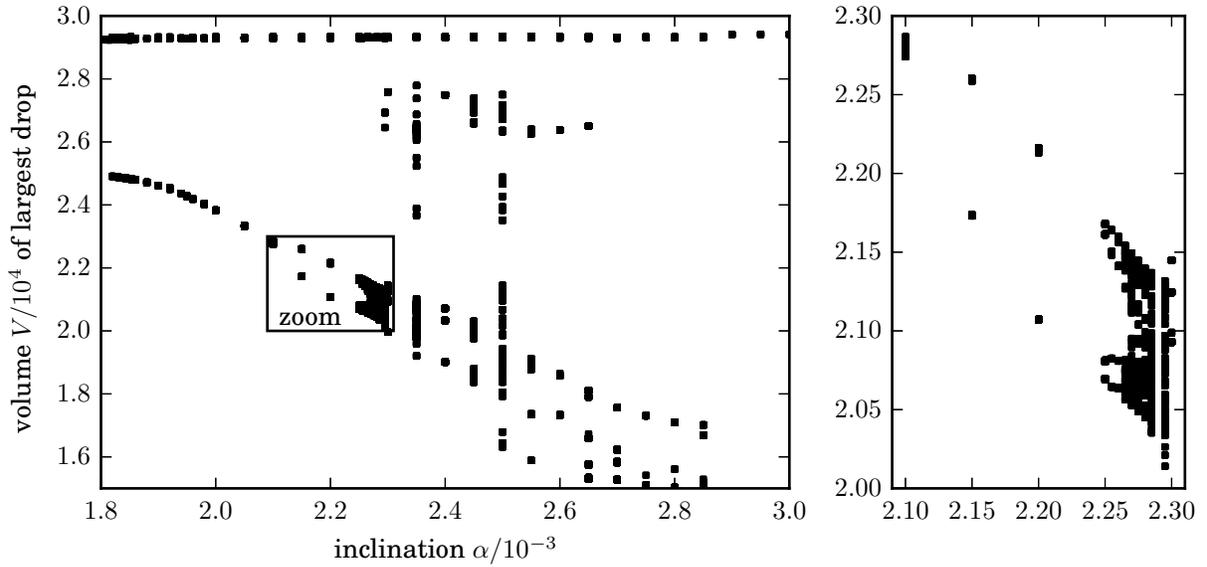}
  \caption{Volumes of all the largest drops that occur in the course of a time simulation at fixed inclination angle $\alpha$ are plotted  as a function of $\alpha$, e.g., for $\alpha = 2.2\times 10^{-3}$ points mark the values indicated by the red bars at the right hand end of the bottom panel of Fig.~\ref{fig:fig5_pearling_volumes_timeseries}.}
  \label{fig:fig_pearling_volumes}
\end{figure}

Figure~\ref{fig:fig_pearling_volumes} shows all these occurring volumes of the respective larger drop for different inclination angles, i.e., at $\alpha = 2.2 \times 10^{-3}$ in Fig.~\ref{fig:fig_pearling_volumes} the shown values correspond to the upper three symbols on the right border of Fig.~\ref{fig:fig5_pearling_volumes_timeseries}.
The period doubling is visible as the transition at $\alpha \approx 2.15 \times 10^{-3}$ from two to three occurring values for the volume, i.e.,
there is always one value corresponding to the volume of the coalesced drop and there are one and two values of split drops before and after the period doubling, respectively. Further increasing the inclination angle, at $\alpha \approx 2.25 \times 10^{-3}$ another period doubling occurs, where the two volumes that occur after pearling again each split up into two slightly different values. The resulting time-periodic state corresponds to a period-4 cycle. These two period doublings are the beginning of the classical period doubling route to chaos \cite{Strogatz1994}. The resulting completely irregular behavior can be seen for values such as $\alpha \approx 2.35 \times 10^{-3}$, where a whole range of volumes occurs in the course of one simulation. Typical for this route to chaos is the emergence of periodic windows; i.e., by increasing the inclination angle from a value that shows irregular behavior, one again finds periodic cycles. In this case, they can also represent a situation where after pearling the larger drop is still unstable and emits a second satellite drop before it merges with the first satellite drop again. Therefore there exist time spans during a cycle with three distinct drops.\par
For inclination angles above $\alpha \approx 2.9 \times 10^{-3}$, the drops become long enough to span the whole domain; i.e., they form rivulets and then no pearling or coalescence occurs. But still, these rivulets can exhibit a time-periodic behavior, e.g., its local height varies along its length and these surface-wave-like modulations travel downward. Further details of the behavior of the rivulets are not in our present scope. Note that the inclination at which the domain-spanning rivulets form depends, of course, on domain size. Another option that we do not pursue here further, is to consider not one drop on a domain of length $L_x$, but two identical drops  on a domain of size $2L_x$. They may show a different pearling behavior, because the volume conservation now holds for the two drops and their satellite drops together and not individually for each drop and its own satellite drop. Therefore also other time-periodic behaviors are possible, like two drops that share one satellite drop. 

\subsection{Unstable stationary solutions: Sub-branches II and III}
\label{sec:branchII_III}

\begin{figure}[htb]
  \center
  \includegraphics[width=1.\textwidth]{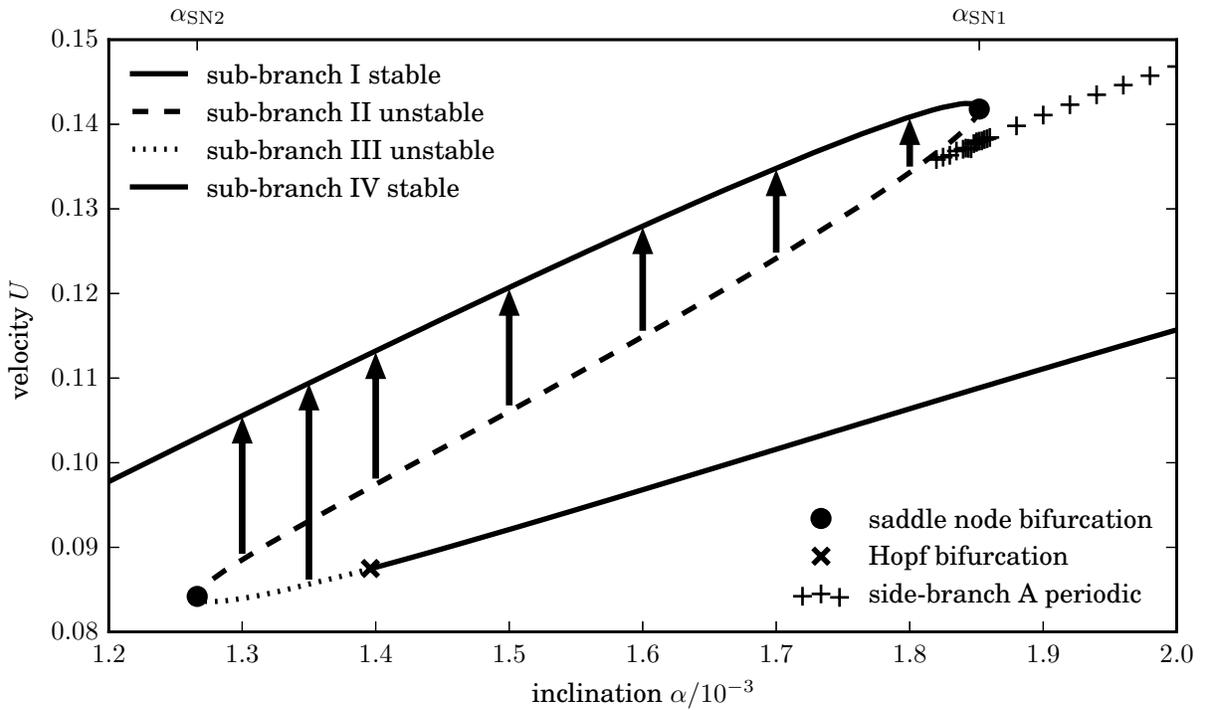}
  \caption{Closeup of Fig.~\ref{fig:fig1_bifurcation} focusing on the monotonously linearly unstable sub-branch~II and the oscillatory linearly unstable sub-branch~III, parts of the neighboring sub-branches~I and~IV and the beginning of side-branch A. The arrows indicate the dynamics that occurs when unstable drops on sub-branches~II and III are disturbed. Normally, they all relax to the stable sub-branch~I (cf. Fig.~\ref{fig:fig9_breathing}).}
  \label{fig:fig7_bifurcationzoom}
\end{figure}  

Next we give some more details on sub-branch~II, i.e., the part of the bifurcation curve in Fig.~\ref{fig:fig1_bifurcation} that connects the first two saddle-node bifurcations. We start at the first saddle-node bifurcation at $\alpha_\mathrm{SN1}$
and follow sub-branch~II with decreasing inclination angle; cf.~Fig.~\ref{fig:fig7_bifurcationzoom} for a zoom of the bifurcation diagram in this region. One clearly sees that side branch~A (discussed in Sec.~\ref{sec:branchA}) emerges from sub-branch~II at an $\alpha$ that is slightly smaller than $\alpha_\mathrm{SN1}$. This point corresponds to the homoclinic bifurcation discussed in Sec.~\ref{sec:branchA}.
All solutions on sub-branch~II are linearly unstable drop solutions that along the branch undergo a morphological change. Namely they develop a protrusion at their back which increases in length with decreasing $\alpha$. This is shown in the snapshots~III and~IV in Fig.~\ref{fig:fig1_bifurcation}. When performing time simulations using these drops as initial conditions, their instability always leads to a dynamical change of the drop shape and velocity. Normally, the protrusion at the end of the drop shrinks until the drop has the shape of the oval drop on the stable sub-branch~I at identical $\alpha$. Alternatively, the drop may undergo a single pearling-coalescence cycle before approaching sub-branch~I.

\begin{figure}[htb]
  \center
  \includegraphics[width=1.\textwidth]{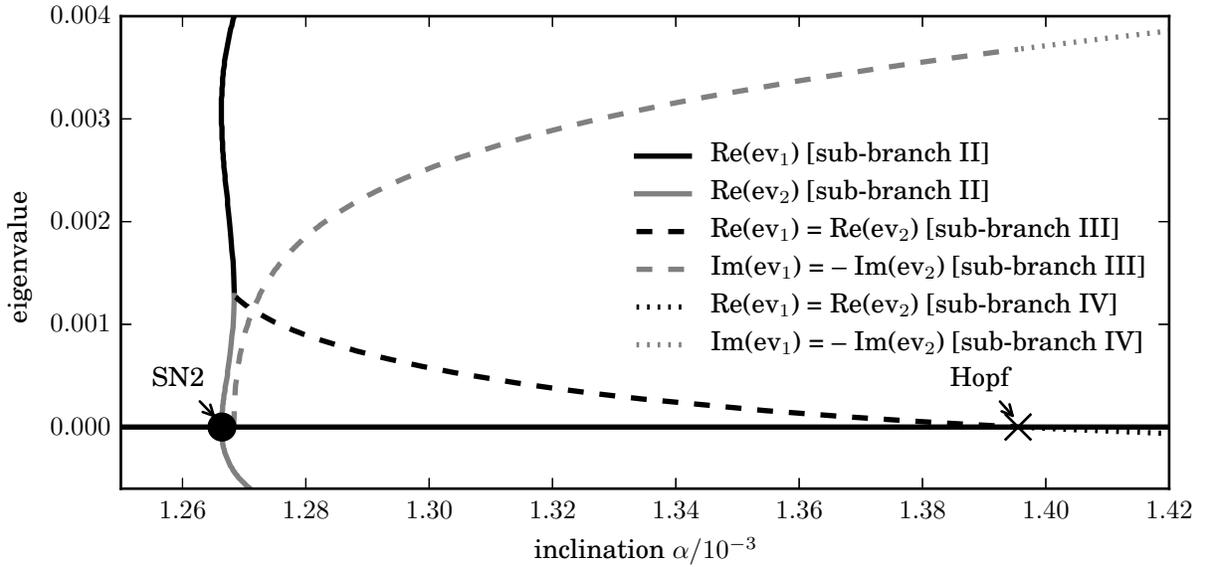}
  \caption{The two leading eigenvalues of stationary drop solutions in the region close to the second saddle-node bifurcation connecting sub-branches~II and~III and the Hopf bifurcation where sub-branch~IV starts. For details see main text.}
  \label{fig:fig8_eigenvalues}
\end{figure}

At the inclination $\alpha_\mathrm{SN2} \approx 1.266 \times 10^{-3}$, a second saddle-node bifurcation occurs in which one could expect the unstable eigenvalue to become stable again. This is, however, not the case illustrated in Fig.~\ref{fig:fig8_eigenvalues} that gives the leading two eigenvalues along sub-branches~II, III, and IV close to $\alpha_\mathrm{SN2}$. One notices that when decreasing $\alpha$ on sub-branch~II close to $\alpha_\mathrm{SN2}$, the magnitude of the first unstable eigenvalue (black solid line) decreases. At the bifurcation at $\alpha_\mathrm{SN2}$ where sub-branch III begins (dashed lines) another real eigenvalue (gray solid line) destabilizes. The two real positive eigenvalues meet at $\alpha = 1.269 \times 10^{-3}$ on the real axis and separate again to form a complex conjugate pair (the black dashed line shows its real part; the imaginary part is shown as the gray dashed line). By increasing $\alpha$ further, the unstable pair crosses the imaginary axis, i.e., it stabilizes in a Hopf bifurcation at $\alpha_\mathrm{hopf} = 1.392 \times 10^{-3}$. There the stable sub-branch~IV starts.

\begin{figure}[htb]
  \center
  \includegraphics[width=1.\textwidth]{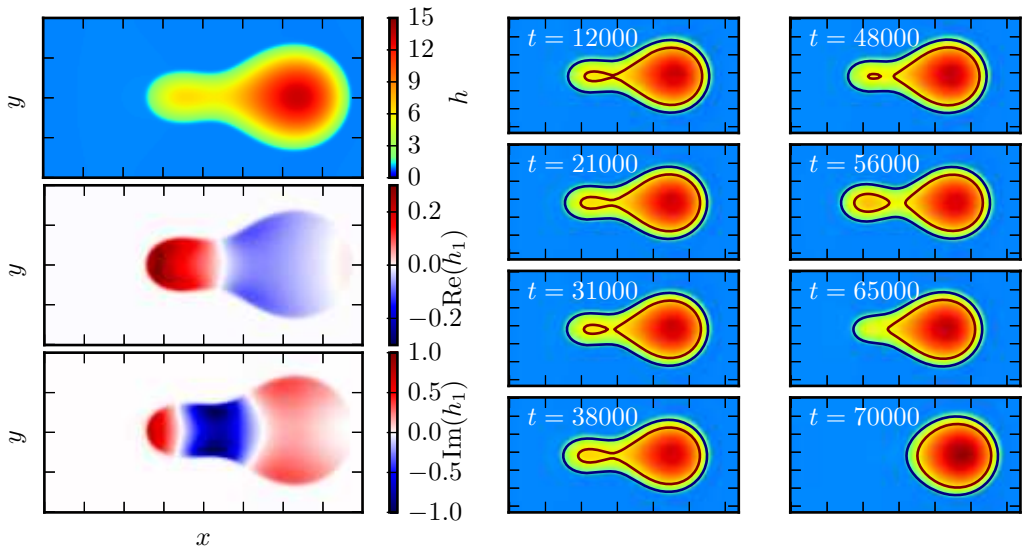}
  \caption{Left panel: Drop profile (top), real (middle), and imaginary (bottom) parts of the complex unstable eigenfunction on sub-branch~III at $\alpha = 1.35 \times 10^{-3}$. Right panel: Time evolution of the droplet in the top left panel. Note the contour lines on the tail of the drop, where a tail oscillation can be observed in the repeated swelling and shrinking. A video of this simulation is available as movie 03 of the Supplementary Material.}
  \label{fig:fig9_breathing}
\end{figure}

\begin{figure}[htbp]
  \center
  \includegraphics[width=1.\textwidth]{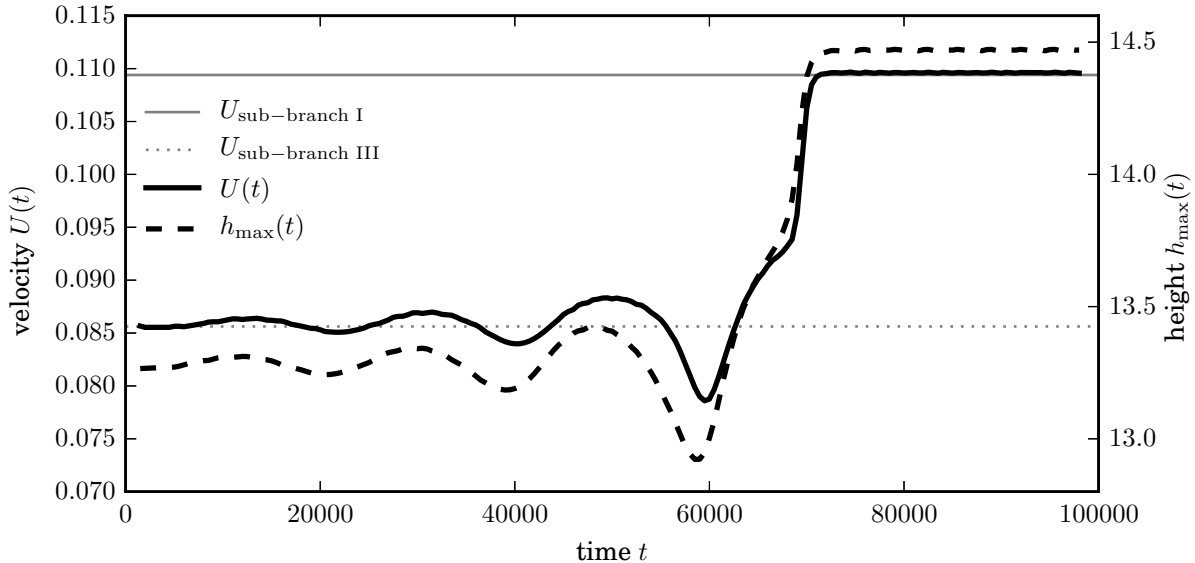}
  \caption{Dependence of drop height (dashed line) and drop velocity (heavy solid line) on time for the unstable oscillatory evolution illustrated by the snapshots in Fig.~\ref{fig:fig9_breathing}. The velocity of the drop oscillates with increasing amplitude around the velocity of the unstable stationary drop from sub-branch III that provided the initial condition  (indicated by the thin dotted horizontal line) until the deformation of the drop is large enough to relax to a drop on sub-branch I (velocity indicated by the thin solid horizontal line).}
  \label{fig:fig10_breathing_velocity}
\end{figure}

Where the leading eigenvalue is complex, the destabilization dynamics of the unstable drops on sub-branch~III is oscillatory, and it exhibits an oscillatory \textsl{breathing}-like behavior of increasing amplitude. Figure \ref{fig:fig9_breathing} illustrates this time evolution as well as the corresponding unstable eigenmode. In the course of the time evolution, the tail of the drop oscillates with an increasing amplitude. When a certain amplitude is reached, the tail gets either absorbed into the main part of the drop, or the tail pinches off as a satellite drop and coalesces with the main drop after one revolution in the periodic domain. In both cases, the final state is a stable solution at identical $\alpha$ on the stable sub-branch~I. This oscillatory behavior is well illustrated in Fig.~\ref{fig:fig10_breathing_velocity}, which shows the dependence of drop height and drop velocity on time. After a phase of about two periods ($t \approx 50000$) during which a harmonic oscillation grows, the evolution becomes clearly non linear, i.e., non harmonic. After one more period the drop closely approaches the shape and velocity of a drop on the stable sub-branch~I.

\subsection{Stationary stable solutions: Sub-branch IV} 
\label{sec:branchIV}

Having discussed the unstable sub-branches~II and~III, we move on to the linearly stable sub-branch~IV that starts at the Hopf bifurcation and continues towards larger $\alpha$; cf.~Figs.~\ref{fig:fig1_bifurcation} and \ref{fig:fig7_bifurcationzoom}. We are not able to discuss the branch of time-periodic solutions that has to emerge from the Hopf point. As time simulations around $\alpha_\mathrm{hopf}$ do not show any sign of stable time-periodic solutions we assume that the branch bifurcates subcritically. The structure of the eigenvalues shown in Fig.~\ref{fig:fig8_eigenvalues} makes it most likely that the branch ends nearby in a homoclinic bifurcation on the unstable sub-branch~II.

\begin{figure}[htb]
  \center
     \includegraphics[width=1.\textwidth]{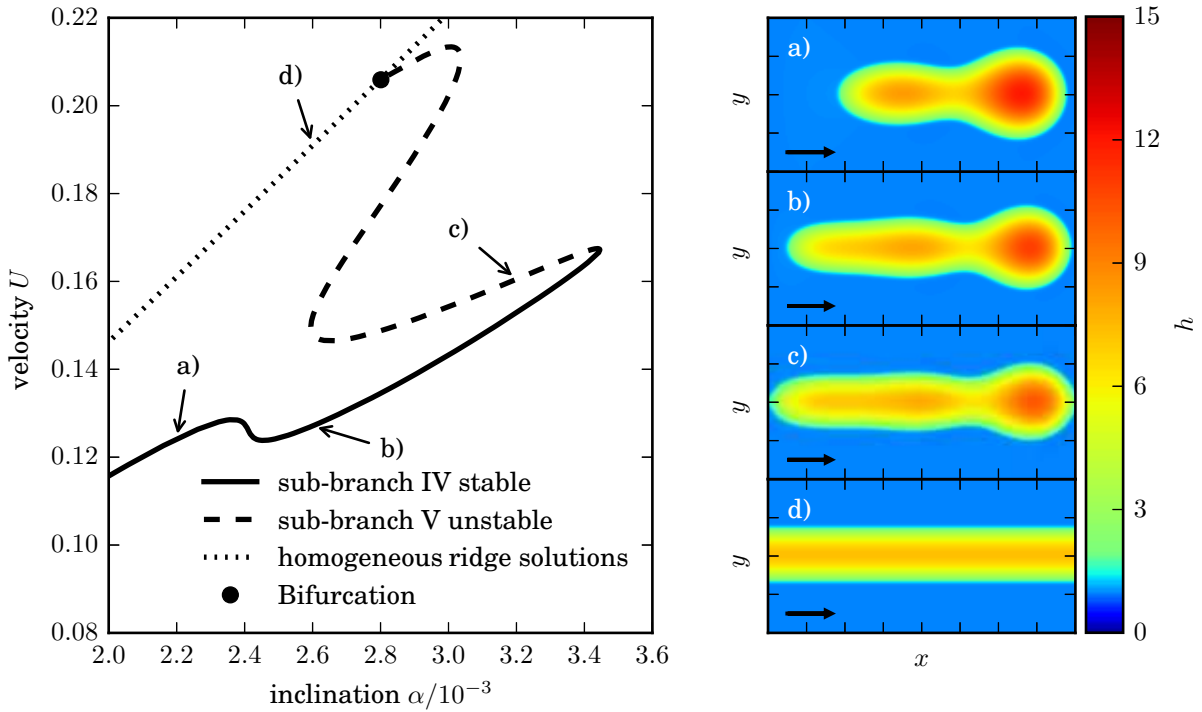}
  \caption{(left) Extension of the bifurcation diagram in Fig.~\ref{fig:fig1_bifurcation} towards larger inclination angles. Along the stable sub-branch IV, a shoulder occurs at $\alpha \approx 2.4\times 10^{-3}$, where the tail of the drop develops another bulge. The drop further elongates, until it spans the whole domain and becomes a translation-invariant rivulet. Panels (a) to (d) give stationary drop profiles at inclinations as indicated in the left panel.}
  \label{fig:fig11_branch4}
\end{figure}

The linearly stable solutions on sub-branch~IV all represent drops with extended protrusions at their rear end and exist in an extended parameter range. As $\alpha$ is increased, the tail gets longer, while the main body of the drop shrinks due to volume conservation [cf.~stationary drops in Figs.~\ref{fig:fig11_branch4}(a) and  \ref{fig:fig11_branch4}(b)]. Note that the existence of this branch of stable non trivial droplets implies that the system shows multistability: The non trivial drops are linearly stable at parameter values where other stable solutions exist: either the oval drops on sub-branch~I or the pearling-coalescence cycles on side-branch~A.

At a certain inclination threshold, the drop length reaches the physical domain size, i.e., its front and back interact [cf. Fig.~\ref{fig:fig11_branch4}(c)]. This results in a number of saddle-node bifurcations as shown in the bifurcation diagram in the left panel of Fig.~\ref{fig:fig11_branch4}. However, here only the first of these at $\alpha \approx 3.5 \times 10^{-3}$ is of interest as it marks the end of sub-branch~IV. The part beyond (sub-branch~V, dashed line in Fig.~\ref{fig:fig11_branch4}) is unstable and non generic as its particular location depends of course on the chosen domain size. The interaction and subsequent fusion of drop front and back results in a rivulet solution that is first modulated but becomes translationary invariant in the $x$ direction [cf. Fig.~\ref{fig:fig11_branch4}(d)] in the final bifurcation (filled circle in Fig.~\ref{fig:fig11_branch4}) that corresponds to a Hopf bifurcation. Note that for a perfect rivulet no velocity can be defined, therefore the dotted line in Fig.~\ref{fig:fig11_branch4} represents the velocity of small traveling modulations on the rivulet that are picked up by the numerical continuation. 

For larger domain sizes, one needs to go to a larger inclination angle to sufficiently lengthen a drop such that it spans the whole domain and becomes a rivulet. Therefore, solutions as in Fig.~\ref{fig:fig11_branch4}(c) would occur for larger inclination angles. This would extend the bifurcation diagram towards larger inclination angles.

\section{Dependence on Drop Volume: Scaling Laws}
\label{sec:powerlaws}

\begin{figure}[htb]
  \center
     \includegraphics[width=1.\textwidth]{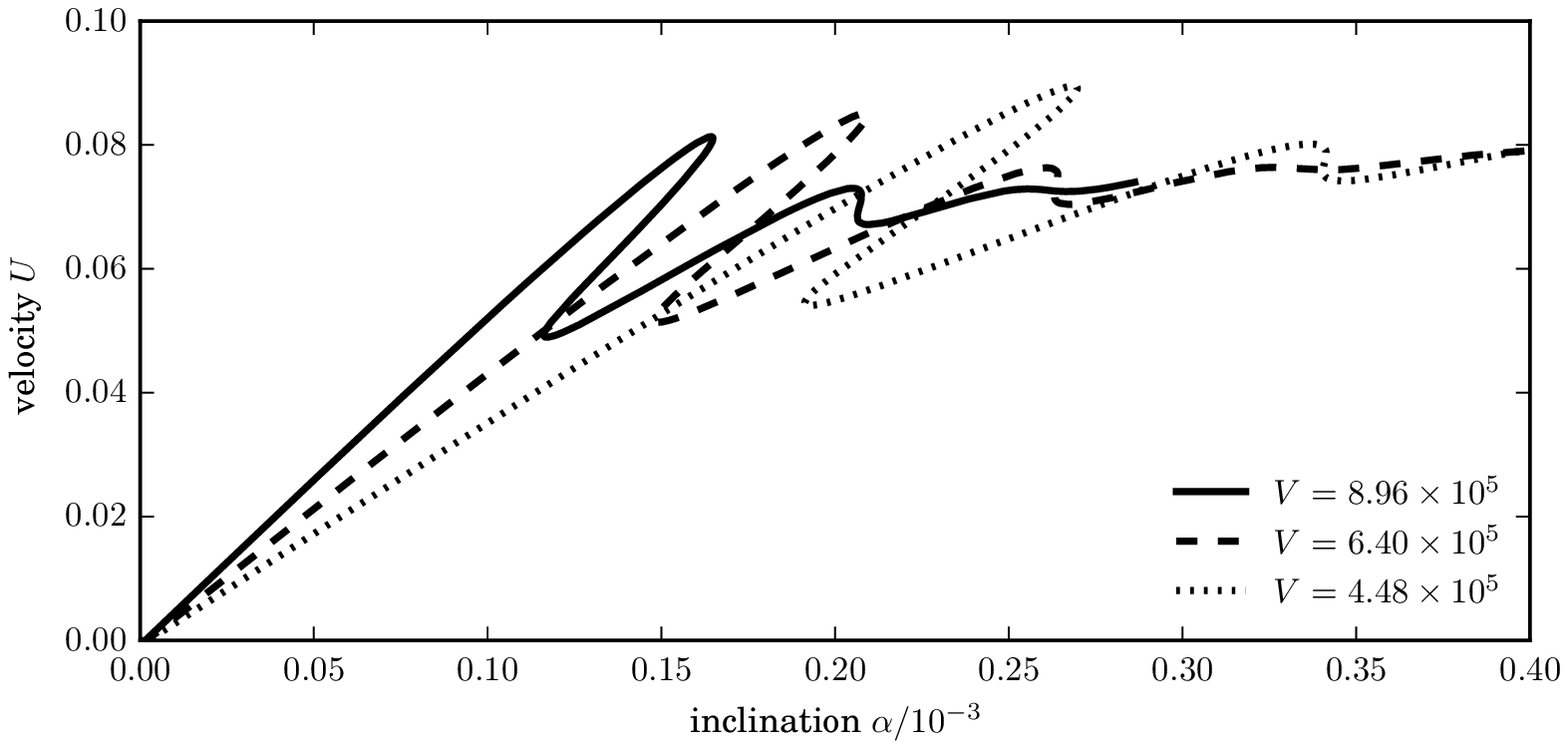}
  \caption{Shown are bifurcation diagrams (drop velocity over inclination angle) for several different drop volumes $V$ as indicated in the legend.
The bifurcation diagram maintains its appearance over the considered range of $V$. With increasing drop size, all characteristic points are shifted towards smaller inclinations (dotted line $\rightarrow$ dashed line $\rightarrow$ solid line). Note that the drop volumes shown here are substantially larger than the case $V=3\times 10^4$ discussed in the previous sections.}
  \label{fig:fig12_volume_dependence}
  \end{figure}

This section focuses on the power laws that have been mentioned in the previous sections. Quantifying these power laws allows us to generalize our observations (that are up to here mainly obtained for a single drop volume) to a broad range of drop volumes. Figure~\ref{fig:fig12_volume_dependence} shows bifurcation diagrams of stationary  drops for several different drop volumes $V$. It shows that its overall appearance does not change over the considered range of drop volumes.

 \begin{figure}[htbp]
  \center
     \includegraphics[width=1.\textwidth]{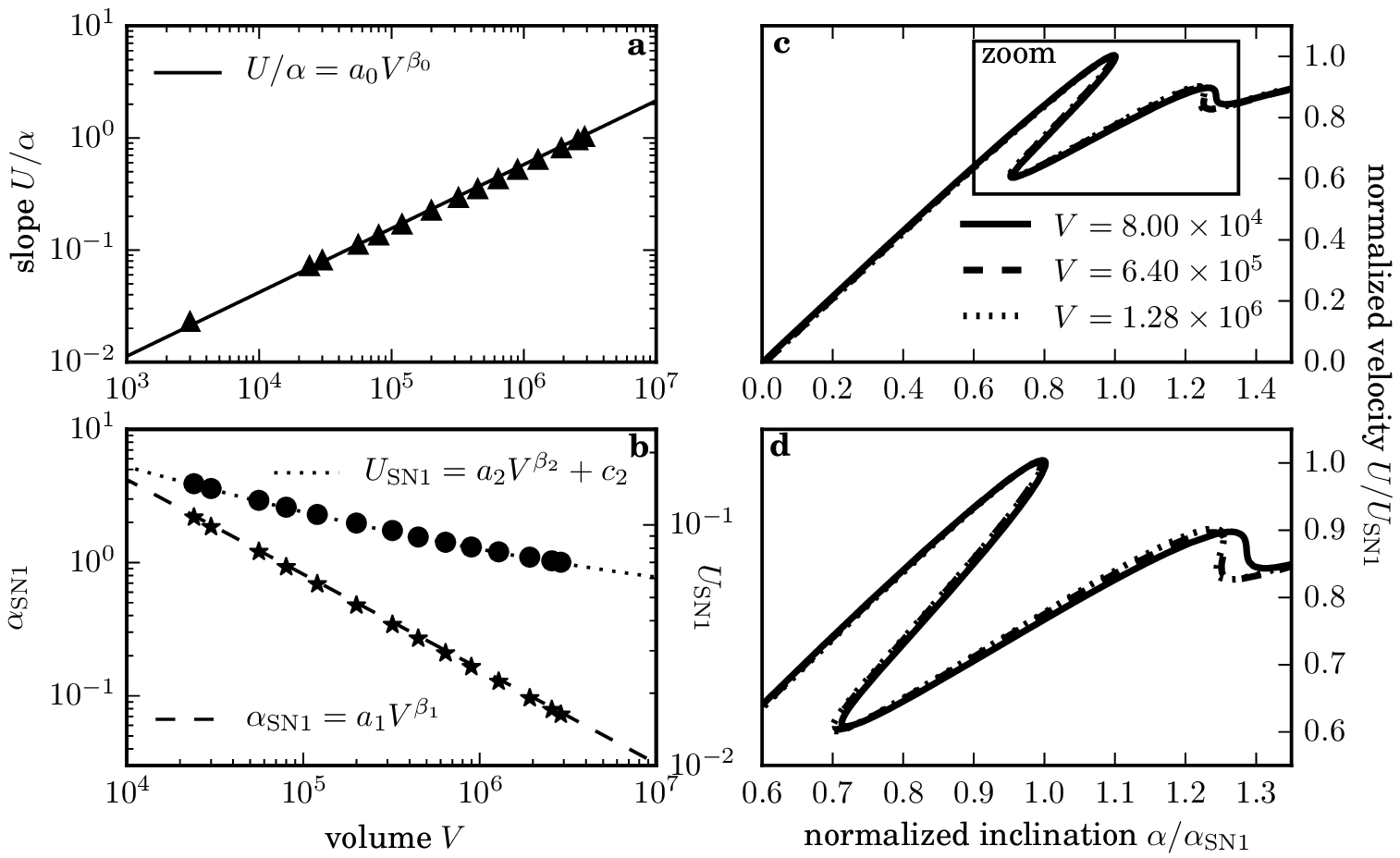}
  \caption{Panel (\textbf{a}) shows the dependence of the slope $U/\alpha$ of the linear part of sub-branch~I in the bifurcation diagram on the drop volume $V$, while panel (\textbf{b}) shows the location of the first saddle-node bifurcation in dependence on $V$. Note that the smallest drop at $V = 3\times10^3$ is missing in panel (\textbf{b}), since the corresponding bifurcation does not exist below a certain volume. Using the power law for the first saddle-node bifurcation to normalize the bifurcation curves at different volume, all graphs collapse onto one master curve, as shown in panel (\textbf{c}) with a closeup in panel (\textbf{d}). This shows the morphological stability of the bifurcation diagrams. Note that the range of drop volumes covered here is much larger than the one shown in Fig.~\ref{fig:fig12_volume_dependence}; therefore the underlying power law is valid for at least three orders of magnitude.}
  \label{fig:fig13_powerlaws}
  \end{figure}

The first part, sub-branch~I (as discussed in Sec.~\ref{sec:branchI}), shows an increasing initial slope as the volume increases. This corresponds to the fact that larger spherical or oval drops slide with higher velocities because the gravitational pulling increases more strongly than the viscous friction (also see Ref.~\cite{XDD2016PF}).  Fitting the linear parts for all analyzed drop volumes $V$ reveals how the slope $U/\alpha$  depends on $V$. These slopes are shown in Fig.~\ref{fig:fig13_powerlaws}(a) for a volume range of about three orders of magnitude. Using a power law ansatz for this dependency we obtain
\begin{align}
\left(\frac{U}{\alpha}\right)_\mathrm{lin}  = a_0V^{\beta_0} \qquad \text{with} \qquad \beta_0 &= 0.569 \pm 0.005 \label{powerlaw_slope} \\
\text{and} \qquad a_0 &= (2.2 \pm 0.2)\times 10^{-4}\,; \notag
\end{align}
i.e., the exponent $\beta_0$ is larger than $1/2$ and smaller than $2/3$. This can be understood from a simple consideration: The velocity $U$ of the sliding drop results from a force balance between gravitational pulling and the viscous friction force. The gravitational force $F_\mathrm{G} $ is proportional to the inclination angle and the volume of the drop, $F_\mathrm{G} \sim \alpha V$. For the friction force $F_\mathrm{F}$ one may discuss two limiting cases: It might either be proportional to the size of the footprint of the drop or to the length of the contact line. In the former case, one expects $F_\mathrm{F} \sim V^{2/3}U$, in the latter case $F_\mathrm{F} \sim V^{1/3}U$. Assuming the force balance $F_\mathrm{G} =F_\mathrm{F}$  holds, one will obtain either $\left(\frac{U}{\alpha}\right)_\mathrm{lin} \sim V^{1-2/3} = V^{1/3}$ or $\left(\frac{U}{\alpha}\right)_\mathrm{lin} \sim V^{1-1/3} = V^{2/3}$.
The numerically obtained exponent $\beta_0 \approx 0.569$ is closer to $2/3$ than to $1/3$. This indicates that the viscous friction is mostly localized close to the contact line and is less related to the footprint of the drop. This is further discussed in Sec.~\ref{sec:dissipation} below, where we analyze the spatial distribution of the viscous dissipation within the drops.\par 

The next power law gives the dependence of the angle of  the first saddle node bifurcation $\alpha_\mathrm{SN1}$ on drop volume $V$ [cf. Fig.~\ref{fig:fig13_powerlaws}(b)]:    
\begin{align}
\alpha_\mathrm{SN1}  = a_1V^{\beta_1} \qquad \text{with} \qquad \beta_1 &= -0.713 \pm 0.005  \label{powerlaw_alpha_bif} \\
\text{and} \qquad a_1 &= (2.9 \pm 0.1) \times 10^{3} \,. \notag
\end{align}
This locus nearly coincides with the angle at which the \textsl{pearling} sets in (see discussion of details in Sec.~\ref{sec:branchA}).
Studying the dependence of the corresponding velocity $U_\mathrm{SN1}$ at this bifurcation on $V$ reveals a slightly more complicated power law shown in Fig.~\ref{fig:fig13_powerlaws}(b) that includes an offset:
   \begin{align}
U_\mathrm{SN1}  = a_2V^{\beta_2} + c_2 \qquad \text{with} \qquad \beta_2 &= -0.26 \pm 0.02 \,, \label{powerlaw_vel_bif} \\
a_2 &= 1.5 \pm 0.1 \notag \\
\text{and} \qquad c_2 &= (3.8 \pm 0.2) \times 10^{-2} \,. \notag
\end{align}
On the one hand, the relations Eqs.~(\ref{powerlaw_alpha_bif}) and~(\ref{powerlaw_vel_bif}) allow one to specify the angle up to which a sliding drop with given volume is stable with respect to pearling. On the other hand, fixing the angle and analyzing drops or drop ensembles with different volumes one can predict which drops are stable or unstable. In addition, the powerlaws can be used to rescale the bifurcations diagrams for different volumes. As can be seen in Figs.~\ref{fig:fig13_powerlaws}(c) and \ref{fig:fig13_powerlaws}(d), the diagrams then almost perfectly collapse onto one single master curve, indicating that the morphology of the bifurcation diagram is stable over many decades of drop volumes.

\section{Energy Dissipation in a Sliding Drop} 
\label{sec:dissipation}

\begin{figure}[htbp]
  \center
     \includegraphics[width=1.\textwidth]{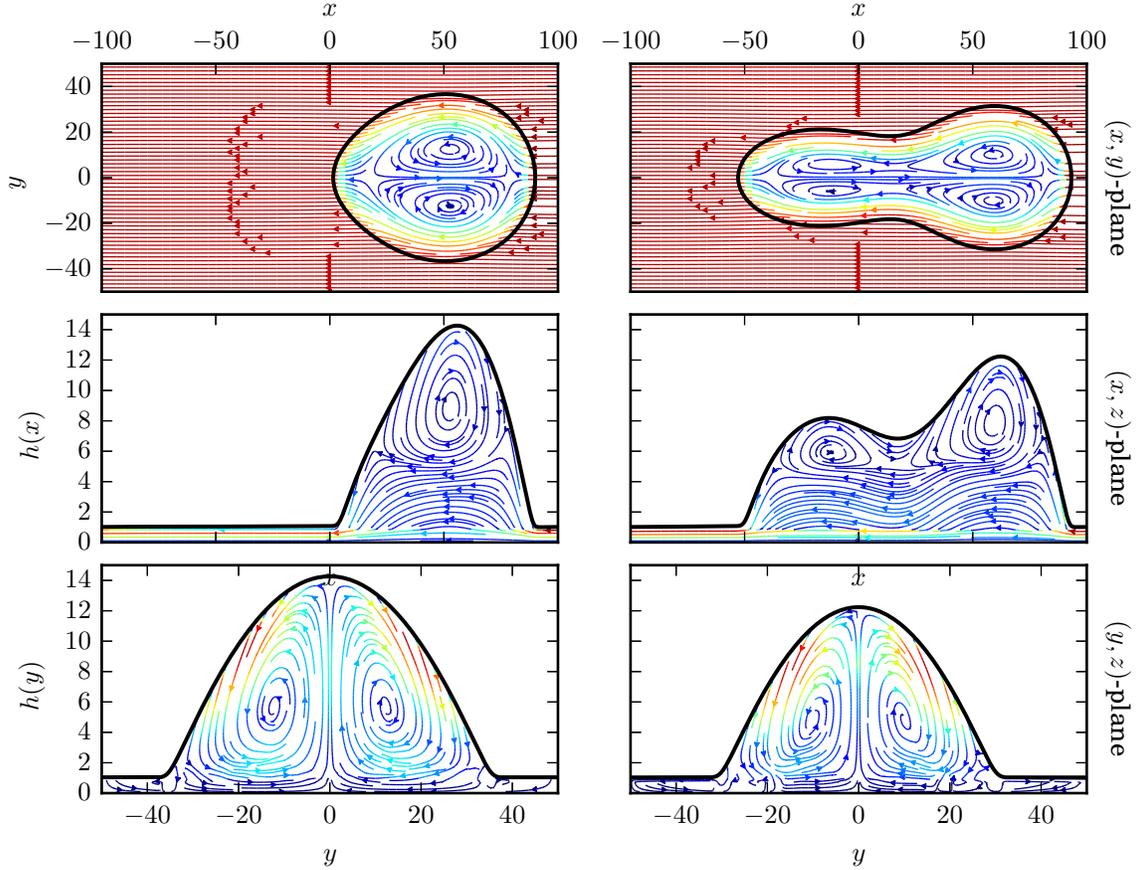}
  \caption{Mean velocity fields in the co-moving frame (determined as described in main text) within particular (left panels) oval cap-like and (right panels) elongated sliding drops representing the stable solution branches~I and~IV, respectively, in Fig.~\ref{fig:fig1_bifurcation}. For both profiles the top, middle, and bottom rows present the top view ($x$-$y$ plane, averaged across the  $z$ direction), side view ($x$-$z$ plane, averaged across the  $y$ direction), and the back-front  view ($y$-$z$ plane, averaged across the $x$ direction), respectively. The colors indicate the absolute value of the respective mean velocity.}
  \label{fig:fig14_velfield}
  \end{figure}

Finally, we investigate how the dissipation mechanisms differ for the various described stationary sliding drops. 
We determine the local and total dissipation for stationary sliding drops based on the dissipation per volume that is in long-wave approximation given by $\eta [\partial_z \mathbf{v}(z)]^2$ where $\mathbf{v}=(u,v)^T$ is the two-dimensional vector of the velocity components in the substrate plane. It depends explicitly on the coordinate $z$; dependencies on the other coordinates are implicit through its dependence on the height profile $h(x,y)$. Employing the no-slip condition at the substrate and force-free conditions at the free surface, in long-wave approximation the $x$ and $y$components of the velocity field in the laboratory frame are given by the parabolic profile \cite{Oron1997Long}
\begin{equation}
   \mathbf{v}(z) = -3\left[\nabla\left[\Delta h + \Pi(h) \right] + (\alpha, 0)^T\right]\left[\frac{z^2}{2} - hz \right]\,.
\end{equation}
Deriving the $z$ component by employing the continuity equation, it is possible to extract the full 3d velocity field from the 2d height profile. Here, in Fig.~\ref{fig:fig14_velfield} the result is presented for two stable drops of different types in the frame moving with the drop. Given are mean velocity fields projected onto certain planes, where we average by integrating over the respective spatial dimension orthogonal to the presentation plane and then pointwise normalizing by the local extent of the drop in this orthogonal direction.

Figure~\ref{fig:fig14_velfield} shows that the velocity of the wetting layer is in the negative $x$ direction and large as compared to the velocity within the drop. This is expected, as it is nearly at rest in the laboratory frame. In consequence, the flow very close to the substrate is laminar within both drops. For the simple oval sliding drop corresponding to solution~I in Fig.~\ref{fig:fig1_bifurcation} (see also Fig.~\ref{fig:fig15_bifurcation_diss} below), two rolls symmetrical with regard to the reflection in the $y$ direction fill the upper part of the drop (left column in Fig.~\ref{fig:fig14_velfield}). For elongated drops (right column in Fig.~\ref{fig:fig14_velfield}), these roll structures split up in the sliding direction, resulting in another pair of rolls within the rear protrusion.

 \begin{figure}[htbp]
  \center
     \includegraphics[width=1.\textwidth]{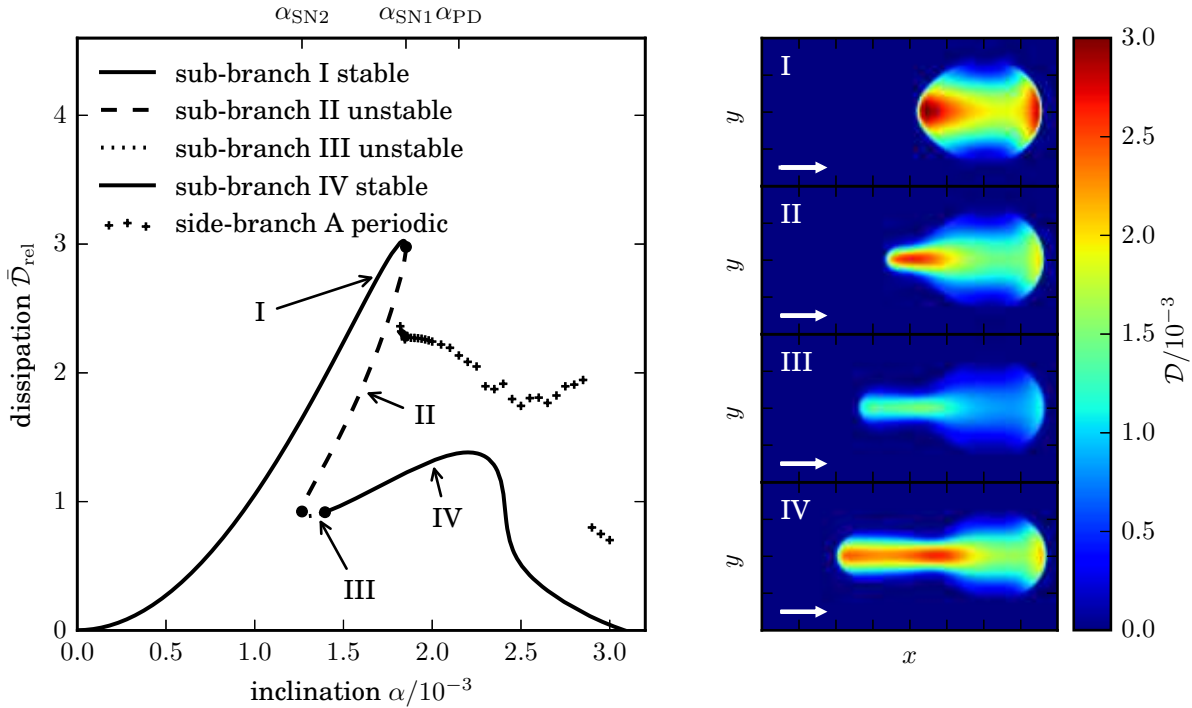}
  \caption{Relative total dissipation $\bar{\mathcal{D}}_\mathrm{rel}$ is given as a function of $\alpha$ for all branches described in Sec.~\ref{sec:results}. The panels on the right hand side show the local per-substrate-area  dissipation $\mathcal{D}(x,y)$ corresponding to the stationary droplet profiles shown on the right hand side of Fig.~\ref{fig:fig1_bifurcation}. The values for the time-periodic solutions obtained by DNS are time-averaged and are therefore only reliable below the onset of the period doubling cascade at $\alpha_\mathrm{PD} \approx 2.15\times10^{-3}$. For the given drop volume the relative dissipation reads $\bar{\mathcal{D}}_\mathrm{rel}[h] \approx \bar{\mathcal{D}}[h] - 69.13\,\alpha^2$.}
  \label{fig:fig15_bifurcation_diss}
  \end{figure}  
 
We now turn to a discussion of the dissipation inside the drop, where the local per-substrate-area dissipation in long-wave approximation is obtained by integration across the film height. In dimensionless form it is
\begin{align}  
    \mathcal{D}(x,y,t) &= \int_0^{h(x,y,t)} \frac{1}{3}\left[\partial_z \mathbf{v}(z)\right]^2 dz \\ 
    &= h^3\left[\nabla\left[\Delta h + \Pi(h) \right] + (\alpha, 0)^T\right]^2\,, \label{loc_diss} 
\end{align}
which is equivalent to
\begin{equation}
\mathcal{D}(x,y,t) =  Q(h) \left(\nabla\frac{\delta \mathcal{F}[h]}{\delta h}\right)^2\,.
\end{equation}
Note that the factor $1/3$ in the dissipation corresponds to a factor of $1/3$, which we absorbed into the scaling of the mobility. The total dissipation is then given by
\begin{equation}
    \bar{\mathcal{D}}(t) = \int_{\Omega} \mathcal{D}(x,y,t)\, dx\,dy \,.
\end{equation}

We use the $x$-translation invariant rivulet $h_\mathrm{r}(y)$ as reference state. It exists for any inclination and is approached by the studied drop solutions at large inclination angles (cf.~Sec.~\ref{sec:branchIV}, Fig.~\ref{fig:fig11_branch4}).
Writing the total dissipation as a functional $\bar{\mathcal{D}}[h]$ we have for the relative total dissipation
\begin{align}
      \bar{\mathcal{D}}_\mathrm{rel}[h] &=  \bar{\mathcal{D}}[h] - \bar{\mathcal{D}}[h_\mathrm{r}] \\
     \text{with} \qquad \bar{\mathcal{D}}[h_\mathrm{r}] &= \alpha^2 L_x \int_{\Omega}h_\mathrm{r}^3\, dy \,,
\end{align}
where we have used that for such rivulets there are no lateral mean flows, i.e., $\partial_{yyy}h+\partial_{y}\Pi(h)=0$.
Note that the reference state is mainly subtracted to ensure readability of the graphical representation. The relative dissipation for droplets on the main branch of Fig.~\ref{fig:fig1_bifurcation} is given as solid and dashed lines in Fig.~\ref{fig:fig15_bifurcation_diss}. On sub-branch~I the relative total dissipation monotonously increases with increasing $\alpha$. Beyond the first saddle-node bifurcation at $\alpha_\mathrm{SN1}$, the dissipation $\bar{\mathcal{D}}_\mathrm{rel}$ decreases along the unstable sub-branch II (with decreasing $\alpha$). Beyond the second saddle-node bifurcation at $\alpha_\mathrm{SN2}$ the dissipation first increases again with increasing $\alpha$ along sub-branches~III and~IV.
It reaches a local maximum at about $\alpha \approx 2.2\times10^{-3}$ before decreasing towards the reference rivulet state. 

\begin{figure}[htbp]
  \center
     \includegraphics[width=1.\textwidth]{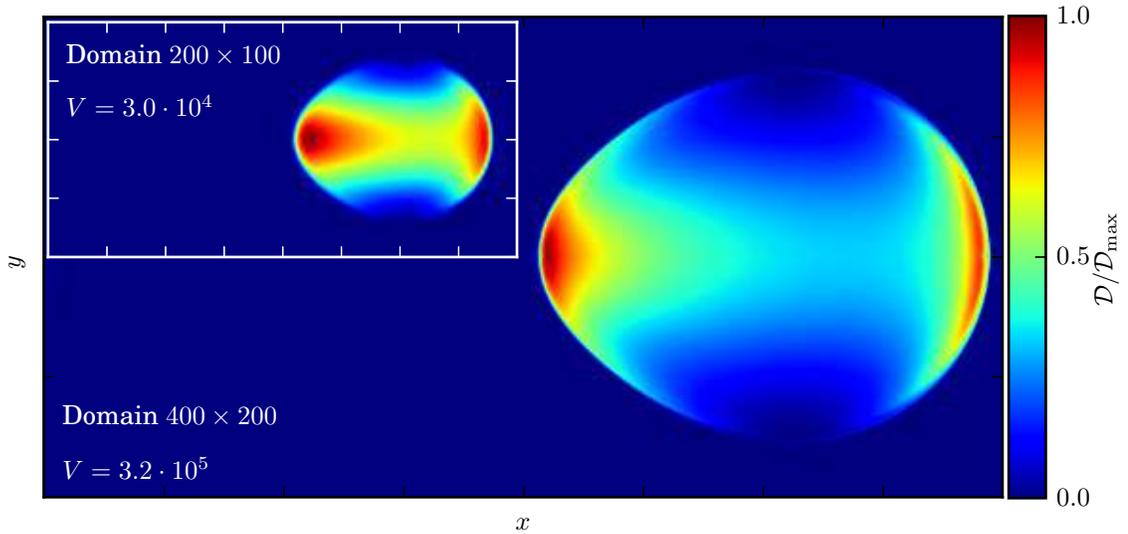}
  \caption{Comparison of the local dissipation of a relatively small drop to a drop of about 10 times larger volume. Both solutions correspond to the region of sub-branch I near the first saddle-node bifurcation. The aspect ratio in the plane is equal in both plots whereas the dissipation is normalized to its respective maximum value of $\mathcal{D}_\mathrm{max} \approx 3.1\times10^{-3}$ for the small drop and $\mathcal{D}_\mathrm{max} \approx 1.4\times10^{-3}$ for the large drop. The local dissipation, i.e., inner velocity gradients, are therefore larger for smaller drop volumes.}
  \label{fig:fig16_big_diss}
  \end{figure}

The spatial distribution of the dissipation within the oval sliding droplets on sub-branch~I is significantly stronger close to the advancing and receding parts of the moving contact line as compared to the central part of the footprint or, indeed, the lateral parts of the contact line. This contrast gets even stronger with increasing droplet volume; cf. Fig. \ref{fig:fig16_big_diss}.
This observation does well agree with the numerically obtained power law Eq.~(\ref{powerlaw_slope}) as we could rationalize it using a simple friction force ansatz [Eq.~(\ref{powerlaw_slope})] if friction at the contact line dominates (see Sec.~\ref{sec:powerlaws}). Similar results for the dissipation are obtained in Ref.~\cite{Kim2002Sliding}, where the dissipation for employed drop shapes and sizes is also assumed to be dominated by viscous effects located at the contact line. Dissipation in 2d spherical cap-like droplets, i.e., ridges with imposed cylinder symmetry are studied in Ref.~\cite{MoVS2011el}. They also find that dissipation at the contact lines dominates.

Interestingly, the dissipation pattern is quite different in the second type of stable sliding drops, i.e., the droplets with elongated tails on sub-branch~IV. They show some dissipation at the leading part of the moving contact line, however, most of the dissipation is localized at the center of the tail along its entire length.  The absolute value of the total relative dissipation is at identical $\alpha$ always larger for the droplets on sub-branch~I than those on sub-branch~IV. The time-averaged dissipation for the time-periodic pearling-coalescence cycles on  side branch~A lies always between the values for sub-branches~I and~IV.
  
\section{Discussion}
\label{sec:concl}

In this work we have presented a theoretical investigation of the bifurcation structure of three-dimensional sliding droplets on a smooth homogeneous inclined substrate. Focusing on small inclination angles and small contact angles and free surface slopes, we have employed an asymptotic long-wave evolution equation for the film thickness profile $h(x,y,t)$ that accounts for capillarity and wettability through Laplace and Derjaguin pressure terms. The latter consists of antagonistic power laws, ensuring that at equilibrium drops of finite equilibrium contact angle coexist with a thin adsorption layer (sometimes called a \textit{precursor film}). The driving by gravity has been incorporated through an additional lateral force term. We have analyzed the model with both a numerical path-continuation technique and direct numerical simulations. The former has allowed us to determine stable and unstable stationary sliding droplets of various kinds as summarized in a comprehensive bifurcation diagram, while the latter has allowed us to determine time-periodic pearling-coalescence cycles of droplets and, in general, to determine the dynamic evolution of the unstable solution types. \par

The bifurcation diagram has been presented in the form of a dependence of the velocity of sliding drops on inclination angle. It shows which droplet types exist at which parameter values, how the various sub-branches are connected, and where stabilities change at the various bifurcations.  Overlaps of linearly stable sub-branches indicate multistability, e.g., between simple stable oval sliding drops and stable sliding drops with a prolonged tail. The different sub-branches are connected by saddle-node and Hopf bifurcations that are responsible for the change in stability along the branch.

Interestingly, the side branch of time-periodic pearling-coalescence cycles does not emerge at one of these local bifurcations but emerges from an intermediate point on an unstable sub-branch in a global homoclinic bifurcation, i.e., without change of linear stability.  The behavior on this sub-branch has been found to correspond to interesting dynamics: first, a repeated cycle of splits into separate drops and their subsequent coalescence. For increasing $\alpha$, these cycles undergo a cascade of period doublings resulting in irregular behavior at larger angles. This closely resembles the behavior found for dripping faucets in dependence of the flow rate \cite{DrHi1991ajp}. One may even argue that our transition between individual oval drops to the stable drops with prolonged tails bears a passing resemblance to the dripping-jetting transition observed for a dripping faucet \cite{ASPB2004prl}. 

We have generalized our results by studying  the influence of the droplet volume on the encountered transitions in behavior. We have found that the general appearance of the bifurcation diagram is universal over several orders of magnitude in volume (also cf.~note \footnote{Note that by translating the results into dimensional quantities as described in the previous footnote, the drops analyzed here would either be physically small or the wetting layer height (or equivalent slip length) would be relatively large; i.e., the scale separation between droplet size and microscopic length is relatively weak. However, the obtained power laws hold over several decades in drop volume and we expect them to also hold for physically large drops. This hypothesis is further supported by the qualitative agreement of our results with the discussed experiments that are all performed with millimetric drops.}). Main features, such as, the volume-dependence of the slope of the linear relation of sliding velocity and inclination angle for simple oval droplets and the dependence of the locus of the first two saddle-node bifurcations can be brought into the forms of power laws. In fact, these power laws are similar to experiment results in Refs.~\cite{Podgorski2001Corners,Kim2002Sliding,legrand2005shape,snoeijer2007cornered}. Here, we found the same functional dependency on the drop volume with an exponent of about $0.57$ whereas the results in the references conclude an exponent of $2/3$.

The time evolution of the droplets in the pearling-coalescence regime can be well compared to earlier numerical approaches to similar models.  Reference~\cite{Schwartz2005Shapes} presented a number of time simulations also using a precursor film model based on a disjoining pressure combining two power laws.  They described the transition to pearling when increasing the driving force and mentioned that a large droplet may decay into an almost chaotic pattern. Reference~\cite{Koh2009} explored the influence of the precursor height, mesh resolution, and numerical accuracy on the morphologies.  Shapes of sliding droplets are also investigated in Ref.~\cite{Pesc2015jcp} employing a slip model and studying droplets of finite support; this allows these authors to observe a spectrum of droplet shapes from ovals to drops with monotonic or non monotonic tails; however, the pearling itself can not be observed. Our results significantly add to these partial pictures by providing the entire bifurcation structure, including stable and unstable droplets and their transitions. Numerical and analytical approaches focusing on the cusp formation \cite{amar2003transition,snoeijer2007cornered} also obtain a universal dependence of the sliding velocity on the inclination angle when staying below the pearling instability. However, asymptotic approaches to the cusp properties do not allow for the identification and further analysis of the pearling instability as a bifurcation, which is what we have presented here.

Another important aspect is our analysis of local and global dissipation in the sliding drops.  Calculating the dissipation by inner velocity gradients of the drop we have found that for the simple oval drops the dissipation is localized in the vicinity of the advancing and receding parts of the moving contact line of the drop.  The first power law with a rather simple friction ansatz for a sliding spherical cap suggests the the friction is localized at the edge of the cap, in total agreement with the spatial distribution of the dissipation as it results from internal friction. Interestingly, the dissipation pattern is quite different in the droplets with elongated tails, where it is localized at the center of the tail along its entire length.

A possible future extension of our present analysis could be a consideration of inhomogeneities, e.g., substrates with regular or random modulations of their topographical or chemical properties. Such systems have already attracted much experimental \cite{Nakajima2013Langmuir,SBAV2014pre} and theoretical \cite{ThKn2006njp,HKRT2007epje,BKHT2011pre} interest that highlighted interesting phenomena, especially concerning the pinning of sliding drops on heterogeneities.
As the multiplicity of steady and traveling states is much larger on heterogeneous substrates, further effort is needed to understand the various competing influences.
Advancing towards real-world applications, such as the movement of water with pesticides on plant leaves \cite{richard2010recent,mayo2015simulating}, such considerations are obviously necessary.

Another important future step in the theoretical analysis of sliding drops is the analysis of large scale dynamics, i.e., the time evolution of large ensembles of droplets where splitting and coalescence events continuously occur. Interestingly, the here-discussed power law for the onset of pearling not only provides the critical inclination angle at fixed drop volume but also a critical volume at a fixed angle.  In this way, it should be possible to employ the information from our single-drop bifurcation study to predict the behavior of ensembles of interacting drops of various volumes. Such a connection of small-scale and large-scale analysis will be pursued elsewhere.

Finally, we emphasize that the approach pursued here of combining path-continuation and direct time-simulation to establish the bifurcation structure for sliding drops on a two-dimensional substrate that is described by a thin-film equation is equally well suited to closely related equations. Examples are (i) the convective Cahn-Hilliard equation describing phase decomposition under the additional influence of an external lateral driving force or faceting dynamics of growing crystals \cite{GNDZ2001prl} and (ii) the Kuramoto-Sivashinsky equation whose variants describe surface waves and deposition patterns \cite{KeNS1990sjam,VoLi2007prb}. We are not aware of any path-continuation study that would elucidate the bifurcation structure of their fully two-dimensional solutions.

\begin{acknowledgments}
This work was partly supported by the Deutsche Forschungsgemeinschaft within the Transregional Collaborative Research Center TRR 61. \par S.E. and M.W. contributed equally to this work.  
\end{acknowledgments}
\newpage
\section*{References}
\bibliographystyle{unsrt}
\bibliography{./literature}

\end{document}